\documentclass[prd,preprint,superscriptaddress,preprintnumbers,eqsecnum,showpacs,nofootinbib,nobibnotes]{revtex4}
\usepackage{amsfonts,amsmath,amssymb,bm,natbib,color}
\usepackage{graphicx} 

\definecolor{prdblue}{rgb}{0.133,0.118,0.498}
\usepackage[colorlinks=true, pdfstartview=FitV, linkcolor=prdblue, citecolor= prdblue, urlcolor=prdblue]{hyperref}

\newcommand{\be}{\begin{equation}}
\newcommand{\bea}{\begin{eqnarray}}
\newcommand{\ee}{\end{equation}}
\newcommand{\eea}{\end{eqnarray}}

\def\s#1{{\scriptscriptstyle #1}}

\def\noeq#1{(\ref{#1})}
\def\1eq#1{Eq.~(\ref{#1})}

\def\2eqs#1#2{Eqs.~(\ref{#1}) and~(\ref{#2})}
\def\3eqs#1#2#3{Eqs.~(\ref{#1}),~(\ref{#2}) and~(\ref{#3})}

\def\fig#1{Fig.~\ref{#1}}

\def\ie{{\it i.e.}, }

\def\n#1{$(#1)$}

\def\g{\widetilde\Gamma^{ \bf np}}

\def\gp{\widetilde\Gamma^{ \bf p}}

\begin{document}

\title{Unified description of seagull cancellations \\ and infrared finiteness of gluon propagators}

\author{A.~C. Aguilar}
\affiliation{University of Campinas - UNICAMP, 
Institute of Physics ``Gleb Wataghin'',
13083-859 Campinas, SP, Brazil}

\author{D. Binosi}
\affiliation{European Centre for Theoretical Studies in Nuclear
Physics and Related Areas (ECT*) and Fondazione Bruno Kessler, \\Villa Tambosi, Strada delle
Tabarelle 286, 
I-38123 Villazzano (TN)  Italy}

\author{C.~T. Figueiredo}
\affiliation{University of Campinas - UNICAMP, 
Institute of Physics ``Gleb Wataghin'',
13083-859 Campinas, SP, Brazil}

\author{J. Papavassiliou}
\affiliation{\mbox{Department of Theoretical Physics and IFIC, 
University of Valencia and CSIC},
E-46100, Valencia, Spain}

\begin{abstract}

We present  a generalized theoretical  framework for dealing  with the important issue  of dynamical mass generation  in Yang-Mills theories, and,  in   particular,  with  the   infrared  finiteness  of   the  gluon propagators, observed  in a  multitude of recent  lattice simulations. Our  analysis is  manifestly  gauge-invariant, in  the  sense that  it preserves   the  transversality   of   the   gluon  self-energy,   and gauge-independent, given  that the  conclusions do  not depend  on the choice  of  the gauge-fixing  parameter  within  the linear  covariant gauges.   The  central construction  relies  crucially  on the  subtle interplay  between  the  Abelian  Ward  identities  satisfied  by  the nonperturbative vertices and a special integral identity that enforces a vast number  of `seagull cancellations' among the  one- and two-loop dressed  diagrams of  the gluon Schwinger-Dyson equation.   The key result of  these considerations is  that the gluon  propagator remains rigorously  massless,  provided  that  the  vertices  do  not  contain (dynamical) massless poles.  When such poles are incorporated into the vertices,  under  the  pivotal  requirement of  respecting  the  gauge symmetry  of  the  theory,  the terms comprising  the  Ward identities  conspire in such a way as to still enforce the total annihilation  of  all quadratic divergences, inducing, at the same time, residual contributions  that account for the saturation of gluon propagators in the deep infrared.

\end{abstract}

\pacs{
12.38.Aw,  
12.38.Lg, 
14.70.Dj 
}

\maketitle

\section{Introduction}                                                      
The infrared finiteness of the gluon propagator, 
namely the fact that its scalar form factor, $\Delta (q^2)$, 
saturates at a finite (nonvanishing) value in the low-energy region,
has been established in the Landau gauge
by means of large-volume lattice simulations both for SU(2)~\cite{Cucchieri:2007md,Cucchieri:2007rg,Cucchieri:2009zt,Cucchieri:2010xr} and SU(3)~\cite{Bogolubsky:2009dc,Bogolubsky:2007ud,Bowman:2007du,Oliveira:2009eh}. 
In addition, recent lattice simulations in the linear covariant ($R_{\xi}$) gauges~\cite{Bicudo:2015rma} reveal  
that this  particular property 
is not special to the Landau gauge ($\xi=0$), given that it persists for  values of 
$\xi$ ranging in the interval $[0,0.5]$. Moreover, the inclusion of a small number of  
dynamical quarks (`unquenching') produces a relative suppression to the   
gluon propagator, but preserves the feature of saturation clearly intact~\cite{Ayala:2012pb}.  
Naturally, these results have attracted particular attention, since they offer a 
valuable opportunity to explore the nonperturbative dynamics of Yang-Mills theories, and even though some of the underlying ideas have a rather long 
history~\cite{Cornwall:1979hz,Cornwall:1981zr,Bernard:1981pg,Bernard:1982my,Donoghue:1983fy,Lavelle:1991ve,
Halzen:1992vd,Philipsen:2001ip,Szczepaniak:2001rg,Aguilar:2001zy,Aguilar:2002tc,Maris:2003vk,Szczepaniak:2003ve,Li:2004te,Aguilar:2004sw},  
various field-theoretic mechanisms that may account for this characteristic behavior
have been considered in the more recent literature~\cite{Aguilar:2006gr,Kondo:2006ih,Braun:2007bx,Epple:2007ut,Binosi:2007pi,Binosi:2008qk,Aguilar:2008xm,Boucaud:2008ky,Dudal:2008sp,Fischer:2008uz,Szczepaniak:2010fe,Watson:2010cn,RodriguezQuintero:2010wy,Campagnari:2010wc,Tissier:2010ts,Pennington:2011xs,Watson:2011kv,Kondo:2011ab,Cloet:2013jya,Siringo:2014lva,Binosi:2014aea}.  

A particular set of physical concepts and formal techniques   
for dealing with this important issue has been 
developed in a series of articles~\cite{Aguilar:2006gr,Binosi:2007pi,Binosi:2008qk}, based on the 
Schwinger-Dyson equations (SDEs)~\cite{Roberts:1994dr} derived within the powerful framework obtained 
from the fusion of the pinch technique (PT)~\cite{Cornwall:1981zr,Cornwall:1989gv,Pilaftsis:1996fh,Binosi:2002ft,Binosi:2003rr,Binosi:2009qm} and background field method (BFM)~\cite{Abbott:1980hw,Abbott:1981ke}.
As has been amply emphasized in the literature cited above,  
one of the most prominent features of the PT-BFM framework are the Abelian 
Slavnov-Taylor identities (STIs) satisfied by the full vertices,
a fact that permits, among other things, the systematic organization 
of the one- and two-loop dressed contributions into manifestly gauge-invariant (transverse) subsets.

The main purpose of the present work is to demonstrate how an elaborate interplay 
between the Ward identities (WIs) and a fundamental nonperturbative cancellation operating at the 
level of the gluon self-energy leads invariably to the 
necessity of introducing massless poles in the vertices of the theory, 
in order to accommodate the aforementioned lattice findings. 
In the four items that follow we summarize   
the logical sequence of the pivotal concepts appearing in this article:

{\bf 1}. To begin with, a sharp distinction between the term `STI' and `WI' must be drawn, which is 
easier established in QED, in terms of the photon-electron vertex, $\Gamma_{\mu}(q,p,p+q)$, and the 
electron propagator, $S(p)$. In this textbook context, what we denominate    
`Abelian STI' is the standard Takahashi identity, \mbox{$q^{\mu}\Gamma_{\mu}(q,p,p+q) = S^{-1}(p+q) - S^{-1}(p)$}, whereas 
the term `Ward identity' refers to the 
relation $\Gamma_{\mu}(0,p,p) = \partial S^{-1}(p)/\partial p^{\mu}$, which may be obtained 
from the Taylor expansion of the Takahashi identity around $q=0$.

{\bf 2}. In the SDE of the gluon propagator appear 
three fully-dressed vertices, namely  
the three-gluon, the gluon-ghost, and the four-gluon vertex. Since we work in the PT-BFM framework, 
these vertices contain one background gluon, namely the one entering into the gluon SDE, 
carrying the momentum $q$ of the gluon propagator; all remaining legs 
are `quantum', and are irrigated by the virtual momenta circulating in the loops 
where these vertices are inserted. Thus, when contracted by 
$q^{\mu}$ from the side of their background leg, they satisfy Abelian STIs, in contradistinction  
to what happens when the contracted leg is quantum, in which case non-Abelian STIs are triggered 
(\ie STIs whose tree-level form is modified by contributions from the  ghost propagator and ghost kernels).
Then, if the form factors comprising the vertices do not contain poles of the type $1/q^2$, then in the limit $q\to 0$
one arrives at expressions that are qualitatively similar to the WI of the photon-electron vertex reported above.

{\bf 3}.  These WIs, in turn, expose a set of crucial cancellations that take place 
between the diagrams belonging to each of the subsets shown in 
Figs.~\ref{fig:a1-a2},~\ref{fig:a3-a4}, and~\ref{fig:a5-a6},
leaving a residual contribution that assumes exactly the form of the `seagull identity', namely 
a special type of integral that  vanishes in any regularization procedure preserving translational invariance,
such as the dimensional regularization. As a consequence, the net effect of the 
combined action between the WIs and the `seagull identity' is the total annihilation of any type of term, 
finite or divergent, that could possibly contribute to  $\Delta^{-1}(0)$. 
   
{\bf 4}. Therefore, in order for the gluon SDE to 
yield $\Delta^{-1}(0) =c$, where $c$ is finite and positive,
one of the conditions imposed in the previous steps must be relaxed.  
Given that the STIs must remain intact, since they are a direct reflection 
of the underlying Becchi-Ruet-Stora-Tyutin (BRST) symmetry of the theory\footnote{We will return to this point shortly.}, 
one must allow for the possibility that some of the vertex form factors contain the aforementioned type of poles\footnote{This general notion dates back to Schwinger's 
seminal observation~\cite{Schwinger:1962tn,Schwinger:1962tp}, according to which  
a gauge boson may acquire a mass, even if the gauge symmetry forbids a mass term 
at the level of the fundamental Lagrangian, provided that its vacuum polarization develops a pole at zero momentum transfer.}. 
In the present work we will not concern ourselves with 
the question of how such poles may be dynamically produced~\cite{Jackiw:1973tr,Jackiw:1973ha,Cornwall:1973ts,Eichten:1974et,Poggio:1974qs,Aguilar:2011xe,Ibanez:2012zk}.
Instead, we will assume their formation, and explain 
how the WIs are rearranged in their presence,    
producing the aforementioned cancellations, but leaving residual terms that 
give rise to the desired effect.

Given that certain variations on the aforementioned concepts have appeared in 
some of the cited literature, it would be useful to briefly highlight the main  
novel aspects of the present approach:

\n{i}  A new form  of the  seagull
identity, \1eq{seagull}, is derived,  which  makes  the   `two-loop  dressed'  analysis  far  more
transparent;  indeed,  in  earlier  works only  the  `one-loop  dressed'
diagrams had been addressed~\cite{Aguilar:2009ke}, while  the cancellations operating at two
loops  had  been neither  identified  nor  implemented. Here, instead, {\it the entire set} 
of dressed diagrams comprising the gluon SDE are treated in a unified way.

\n{ii} While in previous works we have relied on the Abelian STIs 
satisfied by the PT-BFM vertices~\cite{Binosi:2009qm}, 
the main tool employed here are the corresponding WIs, given in \3eqs{BQ2-id}{Bc2-id}{IAbelianSTIBQ3}.
Even though these WIs are contained in the STIs, in the sense 
described in item 1 above, their use allows for 
a more elegant and concise demonstration of some of the most central results.

\n{iii}  Whereas in the past all related demonstrations were carried out in the
Landau gauge, the constructions presented here are valid {\it for any value}
of  the gauge-fixing  parameter, within  the context of the  linear
covariant gauges. 

\n{iv} From the  conceptual point of view, 
the theoretical approach elaborated here 
marks a gradual departure from the strict notion of a momentum-dependent 
gluon mass, first introduced in~\cite{Cornwall:1981zr} and subsequently employed in 
a large number of studies.  
Specifically, in recent articles~\cite{Aguilar:2011ux,Binosi:2012sj,Ibanez:2012zk}, 
the infrared finite gluon propagator was parametrized as $\Delta^{-1}(q^2)=q^2 J(q^2) + m^2(q^2)$,
where $J(q^2)$ plays the role of the `kinetic term' or `wave-function' contribution, 
and $m^2(q^2)$ that of the `effective gluon mass', satisfying the 
crucial condition $m^2(0) > 0$. However,  
in contradistinction to the case of the quark propagator, where the 
Dirac structure of the wave- and mass-functions makes their separation completely 
unambiguous, $J(q^2)$ and  $m^2(q^2)$ are strictly distinguishable 
only at $q=0$, while, away from the origin,  
an arbitrary amount of $J(q^2)$ may be allotted to  $m^2(q^2)$.
Instead, throughout our analysis we have refrained from using   
any such parametrization, treating $\Delta (q^2)$ as a {\it single function},  
and focusing exclusively on its behavior at the origin.

\n{v} The above point affects substantially our understanding of the way in which the    
massless poles enforce the preservation of the corresponding STIs.
Roughly speaking, previously the pole part of the three-gluon vertex 
[left-hand side (l.h.s.) of the STI] would furnish, after 
its contraction with the appropriate momentum,  
the $m^2(q^2)$-components of the gluon propagators appearing 
on the right-hand side (r.h.s.) of the STI~\cite{Aguilar:2011ux,Binosi:2012sj,Ibanez:2012zk}.
Instead, in the current interpretation, the contraction of the pole parts 
furnishes a piece that accounts for the infrared finiteness of the corresponding propagators 
at the origin, but has no a-priori restriction on its form for general $q^2$. 
This novel point of view leads to a considerable reassessment of previous standpoints~\cite{Ibanez:2012zk},
as is clearly reflected in the present treatment of the ghost-gluon vertex, 
which may also possess poles without clashing with the nonperturbative masslessness of the ghost.

\n{vi} Last but not least, we identify a procedure based on the particular form factor content of the 
WIs in the presence of poles, which, at least in principle, may corroborate or falsify the proposed mass-generating mechanism. 

Let us now return to point {\bf 4} of the previous discussion, 
and make some additional clarifications. 
One of the  main advantages of the PT-BFM formalism that we employ is
that the transversality  of the gluon self-energy may  be systematically enforced at
the  level  of  the  corresponding  SDE,  because  the  fully-dressed
vertices entering  in its diagrammatic representation  satisfy special
(Abelian) STIs [see \1eq{BQids}]. In fact, the PT-BFM is the only framework where the 
stronger version of the block-wise transversality is realized [see \1eq{blockwise}].
The gluon transversality remains formally exact 
even when a dynamical gluon mass will be generated, because 
there is no modification whatsoever at the level of the 
original Yang-Mills Lagrangian; the infrared finiteness of the gluon propagator stems 
as a special solution of the same SDEs that were derived at the beginning. 
Therefore, the transition from a massless to a ``massive'' gluon propagator 
introduces no explicit BRST breaking at any step of the way.
A subtle objection may be raised, however: one could argue that artefacts 
may appear that are not directly related to the actual mass generating mechanism that we put forth here, 
but rather originate from the fact that our SDEs are derived
within a Faddeev-Popov-type of gauge-fixing scheme
(namely the  BFM), which offers no a-priori control  on the issue of the  Gribov copies. 
This possibility is  difficult to confirm
or discard within the specific context of the SDEs themselves, because
the information on  any possible Gribov overcounting is dissipated when
the minimization  principle is  implemented during their derivation.  
A reasonable answer may be conjectured  (but not rigorously demonstrated) at least about the gluon transversality,  
by studying what happens to it in the context of 
a formalism such as the Gribov-Zwanziger (GZ) quantization procedure~\cite{Gribov:1977wm,Zwanziger:1989mf},
which maintains the Gribov problem under control.

Within this latter formalism, the impact of the Gribov copies on the nonperturbative physics is reduced 
because the gauge condition is implemented by restricting the functional integral over gauge-field configurations that reside  
within the so-called ``first Gribov region''~\cite{Dudal:2008sp,Zwanziger:1989mf,Dudal:2007cw}. 
This restriction is imposed by adding to the usual (Landau) gauge-fixed Yang-Mills action a nonlocal term, 
known as the ``horizon function''.
This special term is accompanied by a massive parameter $\gamma$, known as the Gribov parameter, whose value is
dynamically determined by a self-consistent condition (horizon condition\footnote{In an alternative approach~\cite{Serreau:2012cg, Serreau:2015yna}, a special average  procedure over  the  Gribov  copies is performed;  
the  resulting local field theory is  equivalent to the massive Curci-Ferrari model~\cite{Curci:1976bt}, and 
the  bare gluon mass  is related to the averaging weight that lifts 
the degeneracy between Gribov copies~\cite{Serreau:2015saa}.}). 
It turns out that the restriction thusly imposed on the space of field configurations 
gives rise to a soft breaking of the BRST symmetry~\cite{Dudal:2009xh,Dudal:2012sb,Cucchieri:2014via}. 
The softness of the breaking follows from the fact that the added term has dimension two in the fields; this, in turn,   
ensures the renormalizability of the theory through suitable STIs\footnote{An exact nonperturbative version of the nilpotent  BRST operator for the Gribov-Zwanziger action in the linear covariant gauges
has been recently introduced in~\cite{Capri:2015ixa,Capri:2016aqq}.}.

Returning to the question of the transversality of the gluon self-energy, 
recently it was proved that the GZ action 
exhibits a modified nonperturbative BRST symmetry~\cite{Capri:2015ixa}. In this framework, 
the usual  BRST transformation corresponds to a symmetry at the perturbative level, which must be adjusted when one approaches the nonperturbative regime.  This extended symmetry  has two main effects: (a) controls the gauge-parameter dependence and 
makes the construction consistent with the gauge invariance, and (b) protects the longitudinal component of the gluon propagator, which is not affected by quantum effects, exactly 
as happens in the  standard Faddeev-Popov quantization scheme; this, in turn,  
is tantamount to the nonperturbative transversality of the gluon self-energy.
We therefore conclude that, 
even though no rigorous analogy 
between two distinct formalism (SDEs and GZ) can be drawn at this stage, it seems reasonable to infer that 
no Gribov-related artifacts will affect the main principles of the present work.


The article is organized as follows: 
Sec.~\ref{sec:gsi} is dedicated to the derivation of a new, compact version of the seagull identity.
In Sec.~\ref{sec:awiotpv} we derive in the context of the PT-BFM framework the WIs satisfied 
by the fundamental vertices of the theory.
Then, in Sec.~\ref{sec:tgseato} we demonstrate that, under the assumption that these vertices
do not contain massless poles, the mixed quantum-background gluon self-energy vanishes at the origin. 
In Sec.~\ref{sec:r} we carry out the renormalization, and show that the conclusions of the previous section 
persist at the level of the renormalized gluon self-energy. 
The main result of this article is presented in Sec.~\ref{sec:etscvwmp}, where it is shown that the inclusion of poles in the vertices 
leads to a subtle distortion 
of the seagull cancellations, giving rise to an infrared finite gluon propagator. 
In Sec.~\ref{sec:num} we derive some additional results, and  
carry out a numerical analysis within a simplified setting, thus offering a 
concrete realization of the formal results obtained previously. 
Finally, in Sec.~\ref{sec:c} we present our discussion and conclusions, and in a short Appendix we report for completeness the relevant Feynman rules. 
Let us end this section by clarifying that  
our analysis is carried out using Feynman rules derived in the Minkowski space, and the transition to the 
Euclidean space, where the formulas of dimensional regularization are valid, is implemented in the last step.
For practical purposes, this transition needs to be made explicit only in arriving at \1eq{a6pert}, and, more importantly,  
in Sec.~\ref{sec:num},  where the numerical analysis is performed.


\section{\label{sec:gsi}Generalized seagull identity}                               

In  this section we derive a more general and compact version of the `seagull identity' 
than the one presented in~\cite{Aguilar:2009ke},
which makes the implementation of the resulting `seagull cancellations'
at the level of the gluon SDE far more transparent and efficient, 
allowing for a unified treatment of one-loop and two-loop dressed diagrams.
In fact, a major advantage of this new formulation, to be exploited in the next section, is that 
it disentangles the action of this identity from the presence of explicit seagull diagrams,
such as $(a_2)$ and $(a_4)$, permitting the treatment of contributions that are 
effectively `seagull-like' (in the sense that they are quadratically divergent), 
but are concealed  inside diagrams of more complicated topology, such as $(a_5)$ and $(a_6)$.

To proceed with the derivation,  
let us introduce the integral measure of dimensional regularization, 
\be
\int_{k}\equiv\frac{\mu^{\epsilon}}{(2\pi)^{d}}\!\int\!\mathrm{d}^d k,
\label{dqd}
\ee
where $d=4-\epsilon$ and $\mu$ is the 't Hooft mass, 
and consider the class of vector functions
\be 
{\cal F}_\mu (k) = f(k^2) k_\mu,
\label{Fvect}
\ee
where, for the time being, $f(k^2)$ is some arbitrary scalar function. 
Since \mbox{${\cal F}_\mu$} is an odd function of $k$, one has immediately that in dimensional regularization 
\be 
\int_k {\cal F}_\mu (k) = 0,
\label{intFvect}
\ee
 
Next, impose on $f(k^2)$ the condition originally introduced by Wilson~\cite{Wilson:1972cf}, namely that, 
as $k^2\rightarrow \infty$, it vanishes rapidly enough so that the integral (in spherical coordinates, with $y=k^2$)
\be 
\int_k f(k^2) = \frac{1}{(4\pi)^{\frac{d}{2}}\Gamma\big(\frac{d}{2}\big)} \int_0^\infty\! \mathrm{d}y\, y^{\frac{d}{2}-1} f(y)
\label{convergence}
\ee
converges for all positive values $d$ below a certain value $d^{*}$. Then, 
the integral is well-defined for any $d$ within $(0,d^{*})$, and can be analytically continued outside this interval\footnote{Additional requirements, such as analyticity of the function $f(k^2)$ at $k^2=0$ may be needed in order to perform the analytic continuation of the integrals. Such cases, however, are not relevant to the present work; for a general discussion, see~\cite{Collins:1984xc}.}.

Observe now that within dimensional regularization (or any other scheme that preserves translational invariance) 
one may shift the argument of the function~\noeq{Fvect} by an arbitrary momentum $q$ without compromising the result~\noeq{intFvect}. 
Then, carrying out a Taylor expansion around $q=0$, and using the result
\begin{align}
	{\cal F}_\mu (q+k) &= {\cal F}_\mu (k) + q^\nu\bigg\lbrace\frac{\partial }{\partial q^\nu}{\cal F}_\mu (q+k)\bigg\rbrace_{q=0} + {\cal O}(q^2)\nonumber \\
	&= {\cal F}_\mu (k) + q^\nu\frac{\partial {\cal F}_\mu (k)}{\partial k^\nu} + {\cal O}(q^2),
\label{TaylFvectq0}
\end{align} 
we obtain
\begin{align}
	q^\nu\int_k\frac{\partial {\cal F}_\mu (k)}{\partial k^\nu} = 0,
	\label{projectq}
\end{align}
since, in agreement with \1eq{intFvect}, if we integrate both sides of the above Taylor expansion, the result must vanish order by order.  

Given that the integral has two free Lorentz indices and no momentum scale, it can only be proportional to the metric tensor $g_{\mu\nu}$; in addition, since $q$ is arbitrary, one concludes that~\1eq{projectq} is realized through the `seagull identity'
\be 
\int_k\! \frac{\partial {\cal F}_\mu (k)}{\partial k^\mu} = 0.
\label{seagull}
\ee

Using finally
\be 
\frac{\partial {\cal F}_\mu(k)}{\partial k^\mu} = 2k^2\frac{\partial f(k^2)}{\partial k^2} + df(k^2)\,,
\label{derFvec}
\ee
we recover the original version of this identity, namely~\cite{Aguilar:2009ke}
\be
\int_k k^2\frac{\partial f(k^2)}{\partial k^2} + \frac{d}{2}\int_k f(k^2) = 0 .
\label{seaold}
\ee

In order to elaborate further on some of the concepts introduced in this section, we consider an explicit example, 
namely the case when $f(k^2)$ is a massive tree-level propagator 
\be \label{fmassprop}
f(k^2) = \frac{1}{k^2-m^2}.
\ee
The following points are then worth mentioning:  

\begin{enumerate}
	\item It is easy to verify using \1eq{convergence} that for  
both integrals appearing in \1eq{seaold} we have that  
$d^{*}=2$, so that they both converge in the interval $(0,2)$.

	\item The validity of \1eq{seaold} may be verified explicitly in the case of \1eq{fmassprop}, by applying the standard integration rules of 
dimensional regularization, namely 
\bea 
\int_k \frac{k^2}{(k^2-m^2)^2} &=& -i(4\pi)^{-\frac{d}{2}}\left(\frac{d}{2}\right)\Gamma\bigg(1-\frac{d}{2}\bigg)  (m^2)^{\frac{d}{2}-1}, \nonumber \\
\int_k \frac{1}{k^2-m^2} &=& -i(4\pi)^{-\frac{d}{2}}\Gamma\bigg(1-\frac{d}{2}\bigg) (m^2)^{\frac{d}{2}-1}.
\label{dimint}
\eea

\item It is clear that if the function $f(k^2)$ were such that the two integrals appearing in \1eq{seaold} would converge for $d=4$  
[for example, $f(k^2) = (k^2-m^2)^{-3}$], 
then its validity could be demonstrated through simple integration by parts, namely   (suppressing the angular contribution) 
\be 
\int_0^\infty\!\mathrm{d}y\,y^{\frac{d}{2}}\frac{\partial f(y)}{\partial y} = y^{\frac{d}{2}}f(y)\big\vert_0^\infty - \frac{d}{2}\int_0^\infty\!\mathrm{d}y\,y^{\frac{d}{2}-1}f(y), 
\label{intparts}
\ee
and dropping the surface term.  For the $f(k^2)$ of \1eq{fmassprop} one may still interpret \1eq{seaold} as a result of an integration by parts, 
where the surface term $\frac{y^{\frac{d}{2}}}{y+m^2 }\big\vert_0^\infty$ can be dropped if $d < d^{*}=2$. 
The fact that one obtains the same value for $d^{*}$ as above, 
suggests an underlying self-consistency of the notions and techniques employed.
\end{enumerate}

We  end this  section by  emphasizing  that the  demonstration of  the
seagull  identity presented  above relies  crucially on  translational
invariance\footnote{We thank M. Lavelle for calling our attention to this important point.},  
which ultimately  permits  the shift  of the  integration variable.  
Therefore, regularization  procedures that  do not  possess
this  property  (such as  the  use  of a  hard  cutoff)  are bound  to
invalidate~\1eq{seagull}. This fact, in turn, may appear to clash with
the numerical treatment  that these equations may  undergo, given that
one  generally introduces  such  cutoffs in  the integration  routines
employed.  However, this  potential  inconsistency can be avoided  by
recognizing  that one  may first  implement 
the seagull cancellations formally, 
encode their implications manifestly into the relevant equations, 
and introduce the cutoffs necessary for their numerical evaluation only at the last step. 

\section{\label{sec:awiotpv}Abelian Ward identities of the PT-BFM vertices}
In this section we derive the basic WIs satisfied by the fully-dressed vertices of the theory.
Given that these vertices appear in the gluon SDE derived within the PT-BFM  framework,  their WIs
are of central importance for the considerations that follow.

\subsection{\label{subsec:gf}General framework}
As has been explained in detail in the related literature, the PT-BFM formalism
provides  a manifestly BRST preserving truncation for the gluon propagator SDE, which ultimately 
governs its nonperturbative dynamics~\cite{Binosi:2007pi,Binosi:2008qk}. Within this framework 
it is particularly expeditious to employ directly the standard BFM procedure, 
and write the gauge field $A^a_\mu$ as the sum of a background ($B^a_\mu$) and a quantum ($Q^a_\mu$) component, \ie \mbox{$A^a_\mu=B^a_\mu+Q^a_\mu$}. 
This splitting introduces a considerable proliferation of Green's function, composed by combinations of $B$ and $Q$ fields.
Thus, for example, in the two-point sector, one has \n{i} 
the conventional gluon propagator, with two $Q$-type gluons ($Q^2$), denoted 
by $\Delta_{\mu\nu}^{ab}(q)$, \n{ii} the mixed  background-quantum propagator, with one $Q$- and one $B$-type gluon ($QB$ or $BQ$), 
denoted by  $\widetilde\Delta_{\mu\nu}^{ab}(q)$, and \n{iii} the background propagator, with two $B$-type gluons, denoted by  $\widehat\Delta_{\mu\nu}^{ab}(q)$.

In what follows we will identify the quantum gauge-fixing parameter 
$\xi_Q$ of the BFM, which appears inside quantum loops, with the corresponding  parameter $\xi$ introduced in the renormalizable 
$R_\xi$ gauges, \ie $\xi_Q =\xi$. Thus, 
the $Q^2$ gluon propagator $\Delta^{ab}_{\mu\nu}(q) = \delta^{ab}\Delta_{\mu\nu}(q)$ is given by 
\be 
\Delta_{\mu\nu}(q) = -i\left[\Delta(q^2)P_{\mu\nu}(q) + \xi\frac{q_\mu q_\nu}{q^4}\right]; \qquad P_{\mu\nu}(q) = g_{\mu\nu}-\frac{q_\mu q_\nu}{q^2},
\label{QQprop}
\ee
with inverse
\begin{align}
	\Delta_{\mu}^{\nu}(q) \Delta_{\nu\rho}^{-1}(q) &= g_{\mu\rho};&\Delta_{\nu\rho}^{-1}(q) &= i\left[\Delta^{-1}(q^2)P_{\nu\rho}(q) + \xi^{-1}q_\nu q_\rho\right].
\label{invQQprop}
\end{align} 
The corresponding self-energy $\Pi_{\mu\nu}(q)$ is transverse, $\Pi_{\mu\nu}(q) = P_{\mu\nu}(q) \Pi (q^2)$, with 
\be
\Delta^{-1}(q^2) = q^2 + i \Pi (q^2).
\label{DeltaPi}
\ee
Completely analogous expressions hold for the $QB$ propagator $\widetilde\Delta_{\mu\nu}^{ab}(q)$, with $\Delta(q^2) \to \widetilde{\Delta}(q^2)$ and $\Pi(q^2) \to \widetilde{\Pi}(q^2)$. It is important to mention that, unlike what happens in QED, all scalar quantities defined above depend also on $\xi$, but this dependence will be suppressed throughout.

Let us finally emphasize that $\Delta(q^2)$, $\widetilde{\Delta}(q^2)$, and  $\widehat{\Delta}(q^2)$ are related by a set of formal, all-order `background-quantum identities'~\cite{Grassi:1999tp, Binosi:2002ez,Binosi:2013cea}, namely
\begin{align}
\Delta(q^2) &= [1 + G(q^2)] \widetilde{\Delta}(q^2) ;&
	\widetilde{\Delta}(q^2) & = [1 + G(q^2)] \widehat{\Delta}(q^2),
\label{propBQI}
\end{align}
with $G(q^2)$ the $g_{\mu\nu}$ component of a special function describing the ghost-gluon dynamics, defined in~\cite{Binosi:2008qk}. 
These identities 
are a consequence of the anti-BRST invariance~\cite{Binosi:2013cea}, and may be generalized to any $n$-point function.

Consider now the SDE that controls the self-energy $\widetilde\Pi_{\mu\nu}(q)$ of the mixed $QB$ propagator. The six diagrams comprising this self-energy are shown in Figs.~\ref{fig:a1-a2},~\ref{fig:a3-a4}, and~\ref{fig:a5-a6}. The fully dressed vertices appearing in the corresponding diagrams, will be denoted by $\widetilde{\Gamma}_{\mu\alpha\beta}$ ($BQ^2$), $\widetilde{\Gamma}_\alpha$ ($B{\bar c}c$), 
and $\widetilde{\Gamma}^{mnrs}_{\mu\alpha\beta\gamma}$ ($BQ^3$); their definition and tree-level expressions are given in Appendix~\ref{app:A}.

When contracted with the momentum carried by the $B$ gluon, these vertices are known to satisfy Abelian STIs; specifically (all momenta entering),
\begin{align}
	q^\mu \widetilde{\Gamma}_{\mu\alpha\beta}(q,r,p) &= i\Delta_{\alpha\beta}^{-1}(r) - i\Delta_{\alpha\beta}^{-1}(p),\nonumber \\
	q^\mu \widetilde{\Gamma}_\mu(q,r,p) &= iD^{-1}(r^2) - iD^{-1}(p^2), \nonumber \\
	q^\mu \widetilde{\Gamma}^{mnrs}_{\mu\alpha\beta\gamma}(q,r,p,t) &= f^{mse}f^{ern} \Gamma_{\alpha\beta\gamma}(r,p,q+t) + f^{mne}f^{esr}\Gamma_{\beta\gamma\alpha}(p,t,q+r) \nonumber \\
	&+ f^{mre}f^{ens} \Gamma_{\gamma\alpha\beta}(t,r,q+p).
	\label{BQids}
\end{align}
where $D(q^2)$ denotes the fully dressed ghost propagator,
and the vertices appearing on the r.h.s.
 of the last equation represents the conventional ($Q^3$) full three-gluon vertices (see Appendix~\ref{app:A} again). 

By virtue of the special Abelian STIs of \1eq{BQids}, it is relatively straightforward to prove the transversality 
of each set of diagrams in Figs.~\ref{fig:a1-a2},~\ref{fig:a3-a4}, and~\ref{fig:a5-a6}~\cite{Aguilar:2006gr}, namely  
\be
q^{\nu} \widetilde{\Pi}_{\mu\nu}^{(i)}(q)= 0,\qquad i=1,2,3.
\label{blockwise}
\ee

\subsection{\label{subsec:wi}Ward identities}
In the ensuing analysis we are interested in the behavior of $\Delta(0)$, or, given the identity~\noeq{propBQI}, $\widetilde{\Delta}(0)$. Thus, the relevant Abelian STIs we need to consider are those obtained by determining the limit of the corresponding identities
in~\1eq{BQids} as the momentum $q$ of the background gluon is taken to vanish. 

The main ingredient that one needs in order to accomplish this task  
is the Taylor expansion of a function $f(q,r,p)$ around $q=0$ (and $p=-r$), given by 
\be 
f(q,r,-r) = f(0,r,-r) + q^\mu\bigg\lbrace\frac{\partial}{\partial q^\mu}f(q,r,p)\bigg\rbrace_{q=0} + {\cal O}(q^2),
\label{Taylf}
\ee
where any possible Lorentz and color structure of the function has been suppressed; evidently, 
after taking the derivative of $f(q,r,-r-q)$ with respect to $q^\mu$ and setting $q=0$, the 
term in curly brackets on the r.h.s. of \1eq{Taylf} becomes a function of $r$ only [of the general form $r_\mu H(r^2)$]. 

In order to fix the ideas, let us first turn to the well-known Abelian model describing the interaction of a photon with a complex scalar field (scalar QED), and consider the Abelian STI (or Takahashi identity) satisfied by the full photon-scalar vertex $\Gamma_\mu(q,r,p)$ [with $\Gamma^{(0)}_\mu(q,r,p)=(p-r)_\mu$],
\be 
q^\mu \Gamma_\mu(q,r,p) = i{\cal D}^{-1}(r^2) - i{\cal D}^{-1}(p^2),
\label{WIAphi2}
\ee
where ${\cal D}(p^2)$ is the fully-dressed propagator of the scalar field [and ${\cal D}^{(0)}(p^2)=i/p^2$]. In order to determine the corresponding WI  we expand both sides of \1eq{WIAphi2} around $q=0$, to obtain
\be 
q^\mu \Gamma_\mu(q,r,p) = q^\mu \Gamma_\mu(0,r,-r) + {\cal O}(q^2) = -iq^\mu \left\{\frac{\partial}{\partial q^\mu}{\cal D}^{-1}((q+r)^2)\right\}_{q=0} + {\cal O}(q^2).
\label{TaylWIAphi2}
\ee
Then, equating the coefficients of the terms linear in $q^\mu$, one obtains the relation
\be
\Gamma_\mu(0,r,-r) = -i\left\{\frac{\partial}{\partial q^\mu}{\cal D}^{-1}((q+r)^2)\right\}_{q=0}=-i\frac{\partial}{\partial r^\mu}{\cal D}^{-1}(r^2),
\label{IWIAphi2q0}
\ee
which is the exact analogue of the familiar textbook WI valid for the photon-electron vertex of spinor QED.

Applying the procedure described above to the first two STIs of~\1eq{BQids}, we obtain the corresponding WIs of the $BQ^2$ and $B\bar c c$, given by 
\begin{align}
\widetilde{\Gamma}_{\mu\alpha\beta}(0,-p,p) &= i \frac{\partial }{\partial p^\mu}\Delta^{-1}_{\alpha\beta}(p);& \widetilde{\Gamma}_{\mu\alpha\beta}(0,r,-r) &= -i \frac{\partial }{\partial r^\mu}\Delta^{-1}_{\alpha\beta}(r),
\label{BQ2-id}	
\end{align}
and, similarly,
\begin{align}
	\widetilde{\Gamma}_\mu(0,-p,p) &= i\frac{\partial }{\partial p^\mu}D^{-1}(p^2);&
	\widetilde{\Gamma}_\mu(0,r,-r) &= -i\frac{\partial }{\partial r^\mu}D^{-1}(r^2).
\label{Bc2-id}
\end{align}

In the case of the $BQ^3$ vertex the derivation of the corresponding WI is slightly more involved. 
Specifically, consider the r.h.s. of the last identity of~\1eq{BQids}; 
then the vertex $\Gamma_{\mu\nu\rho}$ stays the same since it does not depend on $q$, whereas we have
\begin{align}
	\Gamma_{\beta\gamma\alpha}(p,t,q+r) = \Gamma_{\beta\gamma\alpha}(p,-r-p,r) + q^\mu\left\lbrace\frac{\partial}{\partial q^\mu}\Gamma_{\beta\gamma\alpha}(p,t,q+r)\right\rbrace_{q=0} + {\cal O}(q^2),
	\label{BQ3-id}
\end{align}
and similarly for the $\Gamma_{\rho\mu\nu}$ term.

Then, since by  Bose symmetry we have 
\be \label{BoseQ3}
\Gamma_{\alpha\beta\gamma}(r,p,-r-p) = \Gamma_{\beta\gamma\alpha}(p,-r-p,r) = \Gamma_{\gamma\alpha\beta}(-r-p,r,p),
\ee
the zeroth order terms vanish, by virtue of the Jacobi identity:
\be
(f^{mse}f^{ern} + f^{mne}f^{esr}+f^{mre}f^{ens}) \Gamma_{\alpha\beta\gamma}(r,p,-r-p) = 0. 
\ee
Finally, after elementary manipulations, 
the terms linear in $q$ yield
\begin{align} 
\widetilde{\Gamma}^{mnrs}_{\mu\alpha\beta\gamma}(0,r,p,-r-p) &= \left(f^{mne}f^{esr}\frac{\partial}{\partial r^\mu} + f^{mre}f^{ens}\frac{\partial}{\partial p^\mu}\right)\Gamma_{\alpha\beta\gamma}(r,p,-r-p),\nonumber \\
\widetilde{\Gamma}^{mnrs}_{\mu\alpha\beta\gamma}(0,-r,-p,r+p) &= -\left(f^{mne}f^{esr}\frac{\partial}{\partial r^\mu} + f^{mre}f^{ens}\frac{\partial}{\partial p^\mu}\right)\Gamma_{\alpha\beta\gamma}(-r,-p,r+p).
\label{IAbelianSTIBQ3}
\end{align}
\3eqs{BQ2-id}{Bc2-id}{IAbelianSTIBQ3} constitute the central results of this section. To the best of our knowledge these special WIs appear for the first time in the literature.

\subsection{\label{subsec:wiavff}Ward identities and vertex form factors}

Let us finally consider how the above WIs reflect themselves at the level of the form factors that appear in the tensorial decomposition of the corresponding vertices. 

The simplest case is that of the (background) gluon-ghost vertex, which, in terms of the two momenta $q$ and $r$ has the general form 
\be
\widetilde{\Gamma}^{\mu}(q,r,p) = \widetilde{\cal A}_1 (q^2,r^2, p^2) q^{\mu} + \widetilde{\cal A}_2(q^2,r^2,p^2) r^{\mu}. 
\label{ghgen}
\ee
In compliance with the assumptions made when deriving the WIs of \3eqs{BQ2-id}{Bc2-id}{BQ3-id},  
we will postulate for the moment that  
the form factors $\widetilde{\cal A}_1$ and $\widetilde{\cal A}_2$ do not contain poles (kinematic or dynamical) in $q^2$. 
From \1eq{ghgen}, and using the above assumption,  
we obtain immediately that, when $q=0$,  
\be
\widetilde{\Gamma}^{\mu}(0,r,-r)  = \widetilde{\cal A}_2(r^2) r^{\mu}
\label{ghq0}
\ee
and, therefore, from \1eq{Bc2-id} follows directly that 
\be
\widetilde{\cal A}_2(r^2) = -2i \frac{\partial }{\partial r^2}D^{-1}(r^2).
\label{Awi}
\ee
An elementary check of this result is to consider the tree-level vertex at $q=0$, namely (see Appendix~\ref{app:A}) 
\be
\widetilde{\Gamma}_{\mu}^{(0)}(0,r,-r) = -2r_{\mu},
\label{ghtree}
\ee
which is indeed what one obtains from \1eq{Awi} after setting $D^{-1}(r^2) = -ir^2$.

Next, the  tensorial decomposition for $\widetilde{\Gamma}^{\mu\alpha\beta}(q,r,p)$ that is  
suitable for our purposes reads
\be
\widetilde{\Gamma}^{\mu\alpha\beta}(q,r,p) = \sum_{i=1}^{14} \widetilde A_i(q^2,r^2,p^2)\, b_i^{\mu\alpha\beta},
\label{glgen}
\ee
where the basis $b^i_{\mu\alpha\beta}$ is chosen to be
\begin{align}
b_1^{\mu\alpha\beta} &=  q^{\mu} g^{\alpha\beta};&
b_2^{\mu\alpha\beta} &=  q^{\mu} q^{\alpha} q^{\beta};&
b_3^{\mu\alpha\beta} &=  q^{\mu} q^{\alpha} r^{\beta};&
b_4^{\mu\alpha\beta} &= q^{\mu} r^{\alpha} q^{\beta};&
b_5^{\mu\alpha\beta} &=  q^{\mu} r^{\alpha} r^{\beta},
\nonumber\\
b_6^{\mu\alpha\beta} &=  r^{\mu} g^{\alpha\beta};&
b_7^{\mu\alpha\beta} &=  r^{\mu} q^{\alpha} q^{\beta};&
b_8^{\mu\alpha\beta} &= r^{\mu} q^{\alpha} r^{\beta};&
b_9^{\mu\alpha\beta} &= r^{\mu} r^{\alpha} q^{\beta};&
b_{10}^{\mu\alpha\beta} &= r^{\mu} r^{\alpha} r^{\beta},
\nonumber\\
b_{11}^{\mu\alpha\beta} &= q^{\alpha} g^{\beta\mu};&
b_{12}^{\mu\alpha\beta} &= q^{\beta} g^{\alpha\mu};&
b_{13}^{\mu\alpha\beta} &= r^{\alpha} g^{\beta\mu};&
b_{14}^{\mu\alpha\beta} &= r^{\beta} g^{\alpha\mu} . 
\label{thebees}	
\end{align}
As in the previous case, the form factors $\widetilde A_i$ are assumed without poles in $q^2$.
Then, in the limit $q\to 0$, only the components  $b_6$, $b_{10}$, $b_{13}$, and $b_{14}$ survive, so that  
\be
\widetilde{\Gamma}^{\mu\alpha\beta}(0,r,-r) =  \widetilde A_6(r^2) r^{\mu} g^{\alpha\beta} + \widetilde A_{10}(r^2) r^{\mu} r^{\alpha} r^{\beta}
+   \widetilde A_{13}(r^2) r^{\alpha} g^{\beta\mu} +  \widetilde A_{14}(r^2) r^{\beta} g^{\alpha\mu}.
\label{Gq0}
\ee
If we now use the general expression for the gluon inverse propagator given in \1eq{invQQprop} to evaluate the r.h.s of the WI in \1eq{BQ2-id}, and match the resulting tensorial structures with those of \1eq{Gq0}, we obtain  
\begin{align}
	&\widetilde A_6(r^2) = 2 \frac{\partial}{\partial r^2}\Delta^{-1}(r^2);&
\widetilde A_{10}(r^2) &= -2 \frac{\partial }{\partial r^2} \left(\frac{\Delta^{-1}(r^2)}{r^2} \right),\nonumber \\
& \widetilde A_{13}(r^2) =\widetilde A_{14}(r^2) = \xi^{-1} - \frac{\Delta^{-1}(r^2)}{r^2}.
\label{theA}
\end{align}

It is easy to verify the validity of the above results at tree-level, since the $BQ^2$ vertex at $q=0$ reads (see Appendix~\ref{app:A})
\be
\widetilde{\Gamma}_{\mu\alpha\beta}^{(0)}(0,r,-r) =  2 r_{\mu} g_{\alpha\beta} + 
r_{\alpha} g_{\beta\mu} \left(\xi^{-1} - 1 \right) + r_{\beta} g_{\alpha\mu} \left(\xi^{-1}  - 1 \right),
\label{Gtreeq0}
\ee
implying immediately that 
\begin{align}
	\widetilde A_6^{(0)}(r^2) &=2;& \widetilde A_{10}^{(0)}(r^2) &= 0;&\widetilde A_{13}^{(0)}(r^2) &=\widetilde A_{14}^{(0)}(r^2) =  \xi^{-1}  - 1,&
\label{A0}
\end{align}
which is exactly what \1eq{theA} yields upon setting $\Delta^{-1}(r^2) =r^2$. 

The analogous construction for the case of the four-gluon vertex $\widetilde{\Gamma}^{mnrs}_{\mu\alpha\beta\gamma}(q,r,p,t)$
would be particularly cumbersome, given the vast proliferation of tensorial structures appearing in its 
Lorentz decomposition~\cite{Binosi:2014kka,Cyrol:2014kca,Eichmann:2015nra}, and will not be carried out. As we will see in what follows,
although results such as \1eq{Awi} and \1eq{theA} must be used in order to trigger \1eq{seaold} at the one-loop dressed 
level,  the new form of the seagull identity makes no reference to them, and, most importantly, 
obviates the need to dwell on the tensorial decomposition of  $\widetilde{\Gamma}^{mnrs}_{\mu\alpha\beta\gamma}(q,r,p,t)$. 


\section{\label{sec:tgseato} Vanishing of $\widetilde{\Pi}_{\mu\nu}(0)$ in the absence of poles}     

Within dimensional regularization, formulas such as 
\be
\int_{k} \frac{\ln^{n} (k^2/\mu^2) }{k^2} = 0 , \quad n =0, 1,2,\ldots
\label{seapert}
\ee
enforce the masslessness of the gluon to all orders in perturbation theory. The most obvious source of such contributions 
are the so-called seagull diagrams, such as ($a_2$) in \fig{fig:a1-a2}. However, similar contributions are 
concealed inside diagrams that do not have the topological form associated with seagull graphs,
such as ($a_1$) in \fig{fig:a1-a2}, and most notably the two-loop diagrams of \fig{fig:a5-a6}.

The situation becomes significantly more 
complicated nonperturbatively, since there is no mathematical justification in assuming, for example, 
that the equivalent expression of \1eq{seapert} for $n=0$, namely $\displaystyle \int_k \Delta(k^2)$, vanishes. 
But if these contributions are not allowed to vanish individually, they are actually `quadratically' divergent, 
namely they behave as a $\Lambda^2$ in the hard cutoff treatment, or 
as $m^2 (1/\epsilon)$ in dimensional regularization. 
The disposal of such divergences, in turn, would require the inclusion in the original Lagrangian of a
counter-term of the form $m^2 A^2_{\mu}$, which is {\it strictly} forbidden by the gauge ({\it viz.} BRST) invariance.

 In this section we use the Abelian WIs derived above in order to cast the formal expressions that determine
the $g_{\mu\nu}$ component of the gluon self-energy at the origin
into a very particular form. Specifically, we demonstrate that various of the terms that could potentially 
lead to seagull-like contributions cancel against each other, both at one- and at two-loop dressed level, 
and that the remainder vanishes because it triggers precisely the seagull identity of \1eq{seagull}. 
The upshot of all this is that $\widetilde{\Pi}_{\mu\nu}(0)$ is not only finite, and, therefore, no modifications 
to the original Lagrangian are required in the sense described above, but it is, in fact, exactly zero.

In order to forestall possible confusion,
we emphasize that all cancellations among different diagrams identified in this section persist unaltered in the case 
when the main assumption of the absence of poles is relaxed, to be presented in Sec.~\ref{sec:etscvwmp}.
As a result, all quadratically divergent terms cancel against each other as before, and the only crucial difference 
that converts $\widetilde{\Pi}_{\mu\nu}(0)$  from vanishing to finite appears in the last step of implementing \1eq{seagull}.

\subsection{\label{subsec:gencon} General considerations}

The exact (block-wise) transversality of $\widetilde{\Pi}_{\mu\nu}(q)$ guarantees that the form factors 
of $g_{\mu\nu}$ and $q_{\mu} q_{\nu} /q^{2}$ are equal and opposite in sign; therefore, at least in principle, 
one may obtain $\widetilde{\Pi}_{\mu\nu}(0)$ by studying the behavior of 
either one of these two form factors as $q^{2} \to 0$. However, 
the mathematical steps required for reaching the final answer are completely different for both cases; 
in particular, the manipulation of the $g_{\mu\nu}$ cofactor is highly nontrivial, 
requiring full use of the ingredients developed in the previous sections, 
whereas the treatment of the $q_{\mu} q_{\nu} /q^{2}$ counterpart is fairly straightforward.

A typical example of this inequivalence is encountered in the treatment of the expression~\cite{Binosi:2012sj} 
\be
I_{\mu\nu}(q) = \int_k\!k_\mu k_\nu f(k,q),
\label{Imunu}
\ee
where $f(k,q)$ is an arbitrary function that remains finite in the limit $q \to 0$. 
Clearly, 
\be
I_{\mu\nu}(q) = g_{\mu\nu}{A}(q^2) + \frac{q_\mu q_\nu}{q^2} B(q^2)\,,
\ee
and the form factors $A(q^2)$ and $B(q^2)$ are given by 
\be
A(q^2) = \frac1{d-1} \int_k\! \left[ k^2 - \frac{(k\!\cdot\!q)^2}{q^2}\right] f(k,q),\,\,\,
B(q^2)  = -\frac1{d-1} \int_k\! \left[ k^2 - d\frac{(k\!\cdot\!q)^2}{q^2}\right] f(k,q).
\label{justI}
\ee
Then, setting $(q\!\cdot\!k)^2 = q^2 k^2 \cos^2\theta$, and using that, for any function $f(k^2)$ 
\be
\int_k\!\cos^2\theta  f(k^2)=\frac1d\int_k\! f(k^2),
\label{costheta-rel}
\ee
we obtain from \1eq{justI} that, as $q \to 0$, 
\be
A(0) = \frac1{d} \int_k\!k^2 \,f(k^2);\qquad B(0) =0.
\label{I0}
\ee
Evidently, the function $f(k^2)$ may be such that the integral defining $A(0)$ diverges, while, for the same 
function, $B(0)$ vanishes; for example if $f(k^2) = \Delta(k^2)$ or $f(k^2) = D(k^2)$ one obtains  quadratically divergent 
$g_{\mu\nu}$ components. 

It is also clear from the above analysis that $A(0)$ and $B(0)$ may be determined    
through the simpler operation of setting $q=0$ directly in the r.h.s. of \1eq{Imunu},  
\be
I_{\mu\nu}(0) = \int_k\!k_\mu k_\nu f(k^2)  
\label{Imunu0}
\ee
which can be only proportional to $g_{\mu\nu}$. Thus, in the absence of a 
contribution proportional to $q_{\mu} q_{\nu} /q^{2}$, one recovers immediately that $B(0)=0$; the value of $A(0)$ may be obtained 
from \1eq{Imunu0} by simply taking the trace, and coincides with that given in \1eq{I0}

In view of the observations made above, 
the procedure that we will follow is to consider the fully dressed diagrams, $(a_j)_{\mu\nu}(q)$, of each subset 
in Figs.~\ref{fig:a1-a2},~\ref{fig:a3-a4}, and~\ref{fig:a5-a6}, and 
set in them directly $q=0$. 
Since the resulting tensorial structures may be only saturated by $g_{\mu\nu}$, 
the $QB$ self-energy at $q=0$ will simply read (color indices will be suppressed whenever possible)
\begin{align}
	\widetilde{\Pi}_{\mu\nu}(0) &= \widetilde{\Pi}(0)g_{\mu\nu};&
	\widetilde{\Pi}(0) &= d^{-1} \widetilde{\Pi}^{\mu}_\mu(0) =\sum_{i=1}^3\widetilde{\Pi}^{(i)}(0),
\label{photzeroPi}
\end{align} 
where $\widetilde{\Pi}^{(i)}(0)$ are obtained by taking the Lorentz trace of the corresponding diagrams, evaluated at $q=0$.
We will denote any such trace by $a_j (q) \equiv (a_j)^{\mu}_\mu(q)$, and in particular $a_j (0) \equiv (a_j)^{\mu}_\mu(0)$.

\subsection{\label{subsec:oldgd}One-loop dressed gluon diagrams}                   

\begin{figure}[!t]
\includegraphics[scale=0.675]{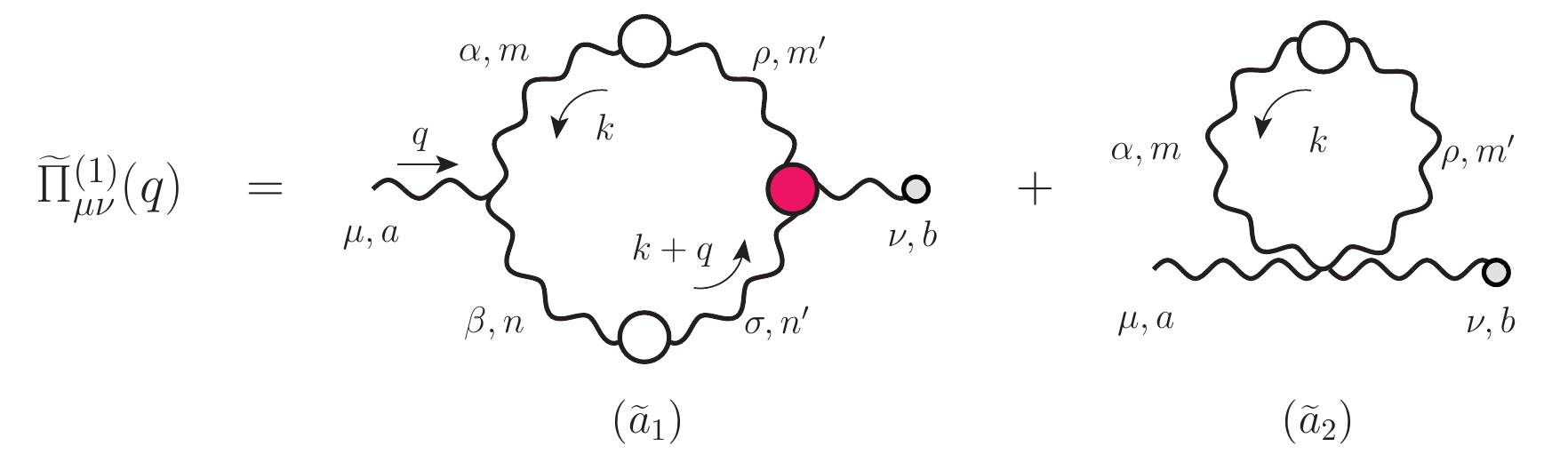} 
\caption{\label{fig:a1-a2}(Color online) One-loop dressed gluon diagrams contributing to the SDE of the $QB$ gluon self-energy. White 
circles indicate fully dressed propagators and the red circle indicates 
a fully dressed three gluon vertex $\widetilde\Gamma$.}
\end{figure}

We start with $\widetilde{\Pi}_{\mu\nu}^{(1)}(q)$, which is given by the sum of the two diagrams shown in \fig{fig:a1-a2}; evidently
\be 
d\widetilde{\Pi}^{(1)}(0) =\widetilde a_1(0) +\widetilde a_2(0),
\label{projQCDzeroPi}
\ee
with 
\bea
\widetilde a_1(0) &=& \frac{1}{2}g^2C_A\int_k \Gamma_{\mu\alpha\beta}^{(0)}(0,k,-k)\Delta^{\alpha\rho}(k)\Delta^{\beta\sigma}(k)\widetilde{\Gamma}^\mu_{\sigma\rho}(0,-k,k)
\label{a1q0}\\
\widetilde a_2(0) &=& -ig^2C_A (d-1) \int_k\Delta^{\alpha}_{\alpha}(k),
\label{a2q0}
\eea
where $C_A$ represents the Casimir eigenvalue of the adjoint representation [$N$ for SU($N$)], and (see again Appendix~\ref{app:A})
\be 
\Gamma_{\mu\alpha\beta}^{(0)}(0,k,-k) = 2k_{\mu}g_{\alpha\beta} - k_{\beta}g_{\alpha\mu} -k_{\alpha}g_{\beta\mu}.
\ee
Using then the WI~\1eq{BQ2-id}, we derive the relation
\be \label{2propsvertex}
\Delta^{\alpha\rho}(k)\Delta^{\beta\sigma}(k)\widetilde{\Gamma}^\mu_{\sigma\rho}(0,-k,k) = -i\frac{\partial }{\partial k_\mu}\Delta^{\alpha\beta}(k),
\ee
so that, after integrating by parts, we may cast \1eq{a1q0} in the form
\be 
\widetilde a_1(0) = -\frac{i}{2}g^2C_A\left\lbrace\int_k \frac{\partial}{\partial k^\mu}\big[\Gamma_{\mu\alpha\beta}^{(0)}(0,k,-k)\Delta^{\alpha\beta}(k)] - \int_k\Delta^{\alpha\beta}(k)\frac{\partial}{\partial k^\mu}\Gamma_{\mu\alpha\beta}^{(0)}(0,k,-k)\right\rbrace.
\label{a1zero}
\ee
Due to the result
\be 
\frac{\partial}{\partial k^\mu} \Gamma_{\mu\alpha\beta}^{(0)}(0,k,-k) = 2(d-1)g_{\alpha\beta},
\label{derQ3tree}
\ee
we then see that the second term of~\1eq{a1zero} cancels exactly against the $\widetilde a_2(0)$ of~\1eq{a2q0}, and we are left with the result
\begin{align}
	d\, \widetilde{\Pi}^{(1)} (0) & =-g^2C_A (d-1)\int_k\frac{\partial}{\partial k_\mu} {\cal F}^{(1)}_\mu(k) ;&
	{\cal F}^{(1)}_\mu(k) &= k_\mu\Delta(k^2).
\label{F1QCD}
\end{align}

\subsection{\label{subsec:oldghd}One-loop dressed ghost diagrams} 
\begin{figure}[!t]
\includegraphics[scale=0.675]{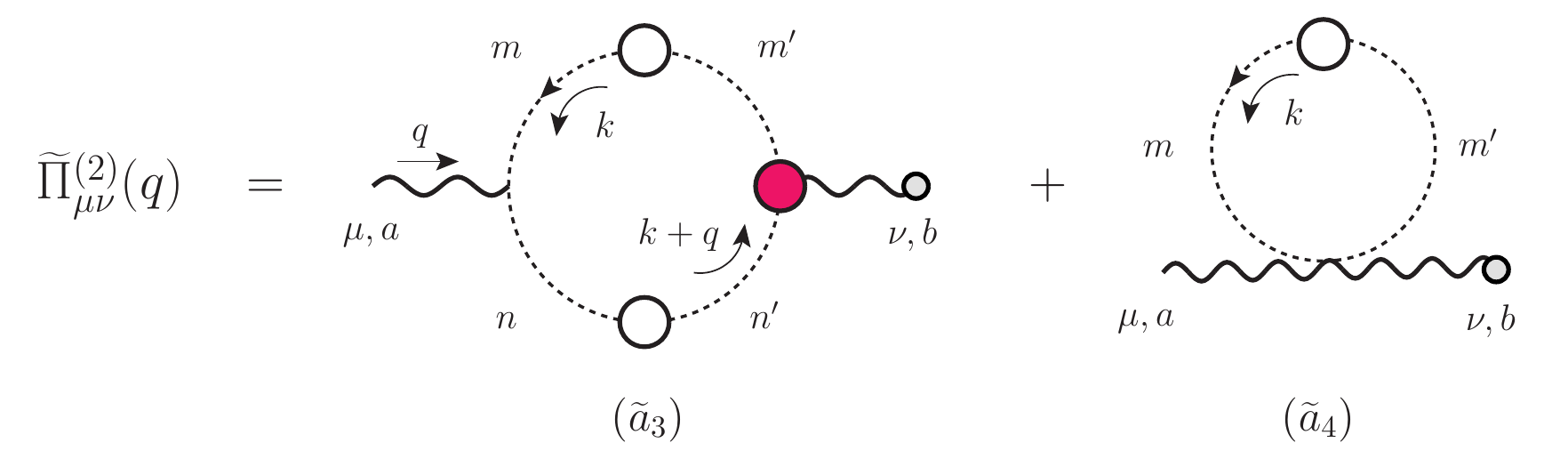} 
\caption{\label{fig:a3-a4}(Color online) One-loop dressed ghost diagrams.}
\end{figure}

Turning to $\widetilde{\Pi}_{\mu\nu}^{(2)}(q)$, the ghost diagrams of \fig{fig:a3-a4} at $q=0$ give 
\begin{align}
	\widetilde a_3(0) &= g^2 C_A\int_k\! k_\mu D^2(k^2)\widetilde{\Gamma}^\mu(0,-k,k), \label{a3} \\
	\widetilde a_4(0) &= -ig^2C_Ad\int_k\! D(k^2). \label{a4}
\end{align} 
Then, from the identity~\noeq{Bc2-id} we obtain 
\be 
D^2(k^2)\widetilde{\Gamma}^\mu(0,-k,k) = -i\frac{\partial D(k^2)}{\partial k^\mu},
	\label{derD}
\ee
and therefore, after integrating by parts, $a_3(0)$ yields
\be 
\widetilde a_3(0) = -ig^2C_A\bigg\lbrace\int_k\frac{\partial}{\partial k^\mu}\big[\Gamma^\mu(0,-k,k) D(k^2)\big] - d\int_k D(k^2)\bigg\rbrace.
\label{a3derD}
\ee
Clearly, the second term in \1eq{a3derD} cancels exactly against $a_4(0)$, 
and we are left with 
\begin{align}
	d\widetilde{\Pi}^{(2)} (0) &= -ig^2C_A\int_k\!\frac{\partial}{\partial k_\mu}{\cal F}^{(2)}_\mu(k);&
	{\cal F}^{(2)}_\mu(k) &= k_\mu D(k^2).
\label{F2QCD}
\end{align}

\subsection{\label{subsec:tldd}Two-loop dressed diagrams}                   

\begin{figure}[!t]
\includegraphics[scale=0.675]{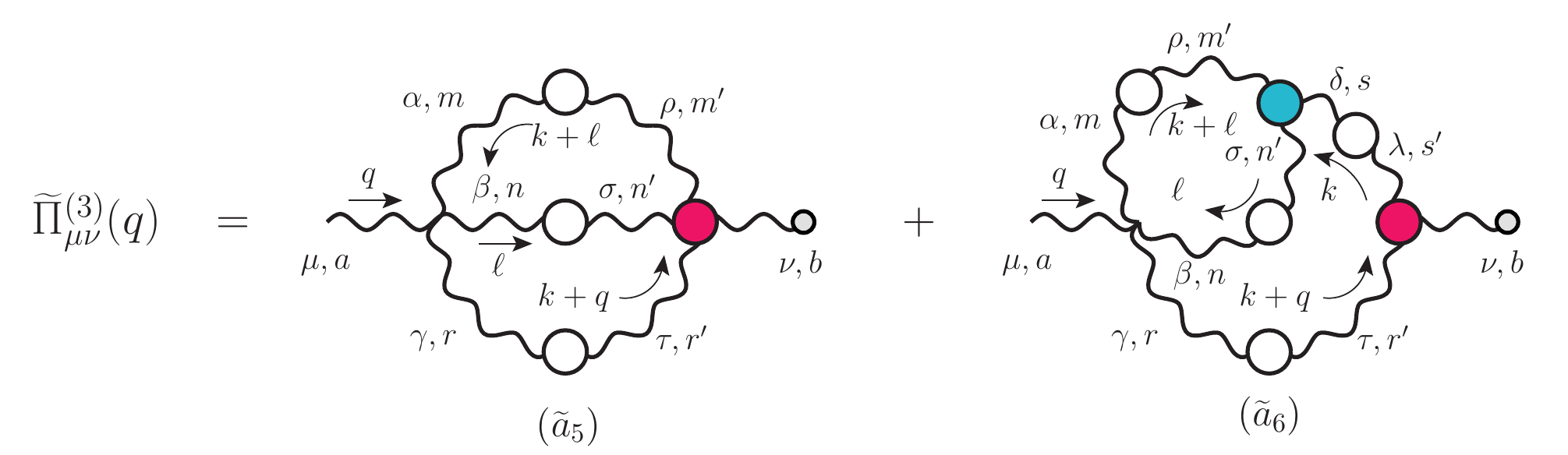} 
\caption{\label{fig:a5-a6}(Color online) Two-loop gluon dressed diagrams.}
\end{figure}

The two-loop dressed gluon self-energy, $\widetilde{\Pi}_{\mu\nu}^{(3)}(q)$, is given by the sum of the two diagrams in \fig{fig:a5-a6}, which
at $q=0$ read
\begin{align}
	\widetilde a_5^{ab}(0) &= -\frac{1}{6}g^4\Gamma_{\mu\alpha\beta\gamma}^{(0)amnr}\int_k\int_\ell\! \Delta^{\alpha\rho}(k+\ell)\Delta^{\beta\sigma}(\ell)\Delta^{\gamma\tau}(k)\widetilde{\Gamma}_{\mu\tau\sigma\rho}^{brnm}(0,-k,-\ell,k+\ell),\nonumber \\
	\label{a5}
	\widetilde a_6(0) &= -i{\cal N}_{\mu\alpha\beta\gamma} \int_k\! Y_\delta^{\alpha\beta}(k)\Delta^{\gamma\tau}(k)\Delta^{\delta\lambda}(k)\widetilde{\Gamma}^\mu_{\tau\lambda}(0,-k,k),
\end{align}
where we have defined
\be 
{\cal N}_{\mu\alpha\beta\gamma} = \frac{3}{4}g^4C_A^2(g_{\mu\alpha}g_{\beta\gamma} - g_{\mu\beta}g_{\alpha\gamma}),
\label{Npre}
\ee
and
\be 
Y_\delta^{\alpha\beta}(k) = \int_\ell\! \Delta^{\alpha\rho}(k+\ell)\Delta^{\beta\sigma}(\ell)\Gamma_{\sigma\rho\delta}(\ell,-k-\ell,k),
\label{Yint}
\ee 
which is proportional to the subdiagram nested inside ($a_6$). One may show that, 
due to the  Bose symmetry of the vertex $\Gamma_{\sigma\rho\delta}$,  $Y_\delta^{\alpha\beta}(k)$ assumes the form~\cite{Binosi:2012sj} 
\be 
Y_\delta^{\alpha\beta}(k) = (k^\alpha g_\delta^\beta - k^\beta g_\delta^\alpha) Y(k^2);\qquad Y(k^2)=\frac1{d-1}\frac{1}{k^2}\,k_\alpha g_\beta^\delta Y_{\delta}^{\alpha\beta}(k).
\label{Yint2}
\ee

Then, using~\1eq{2propsvertex}, we immediately obtain
\be 
\widetilde a_6(0) = -{\cal N}_{\mu\alpha\beta\gamma} \int_k Y_\delta^{\alpha\beta}(k)\frac{\partial \Delta^{\gamma\delta}(k)}{\partial k^\mu}.
\label{a6zero}
\ee
Even though $a_6(0)$ will eventually cancel in its entirety against an analogous contribution from $a_5(0)$, it is instructive to evaluate its form a bit further. Specifically, choosing for simplicity the Landau gauge ($\xi=0$) and using \1eq{Yint2}, we find
\be
\widetilde a_6(0) = \frac{3}{2} i (d-1) g^4C_A^2 \int_k \left[\Delta(k^2) + 2 k^2\Delta'(k^2)\right]Y(k^2).
\label{a6more}
\ee
At this point one may set $\Delta(k^2) \to 1/k^2$ into \1eq{a6more}, and use the 
lowest order perturbative expression for $Y(k^2)$~\cite{Binosi:2012sj},  
\be
Y(k^2)= \frac{-5i} {64\pi^2} \log\!\left(\frac{-k^2}{\mu^2}\right),
\label{Yappr}
\ee
to obtain (in Euclidean space) 
 \be
\widetilde a_6^{\rm pert}(0) = c \int_{k} \frac{ \ln (k^2/\mu^2)}{k^2},
\label{a6pert}
\ee 
with $c$ an irrelevant numerical constant; an exactly analogous expression is obtained for a general value of the gauge-fixing parameter. 
Evidently, \1eq{a6pert} corresponds to the $n=1$ case of \1eq{seapert}, and therefore vanishes.
This, however, is no longer true nonperturbatively, and $a_6(0)$ yields, up to logarithms, a quadratically divergent contribution. 

Turning to the contribution coming from the other two-loop diagram,  after employing the crucial WI of ~\1eq{IAbelianSTIBQ3},  
we find that 
\bea 
\widetilde a_5^{ab}(0) &=& -\frac{1}{6}g^4\Gamma_{\mu\alpha\beta\gamma}^{(0)amnr}\int_k\int_\ell \Delta^{\gamma\tau}(k)\Delta^{\beta\sigma}(\ell)\Delta^{\alpha\rho}(k+\ell) \nonumber \\
&\times& \left(f^{bre}f^{emn}\frac{\partial}{\partial k^\mu} + f^{bne}f^{erm}\frac{\partial}{\partial \ell^\mu}\right)\Gamma_{\tau\sigma\rho}(k,\ell,k+\ell).
\label{a5partialkl}
\eea

Next, we make use of the identities
\bea \label{ffG}
f^{brx}f^{xmn}\Gamma_{\mu\alpha\beta\gamma}^{(0)amnr} &=& \frac{3}{2}C_A^2\delta^{ab}(g_{\mu\alpha}g_{\beta\gamma} - g_{\mu\beta}g_{\alpha\gamma}), \nonumber \\
f^{bnx}f^{xrm}\Gamma_{\mu\alpha\beta\gamma}^{(0)amnr} &=& \frac{3}{2}C_A^2\delta^{ab}(g_{\mu\gamma}g_{\alpha\beta} - g_{\mu\alpha}g_{\beta\gamma}),
\eea
integrate by parts and carry out the appropriate shifts in the integration momenta, to obtain
\be \label{a5Yint}
\widetilde a_5(0) = -{\cal N}^\mu_{\alpha\beta\gamma} \bigg\lbrace \frac{2}{3}\int_k \frac{\partial}{\partial k^\mu}\big[\Delta^{\gamma\delta}(k)Y_\delta^{\alpha\beta}(k)\big] - \int_k Y_\delta^{\alpha\beta}(k)\frac{\partial \Delta^{\gamma\delta}(k)}{\partial k^\mu}\bigg\rbrace,
\ee
where the color factor $\delta^{ab}$ has now been omitted. 
Evidently, the second term on the r.h.s cancels exactly the entire contribution \1eq{a6zero}, as anticipated, 
and we finally obtain 
\begin{align}
	d\widetilde{\Pi}^{(3)} (0) &= i(d-1)g^4C_A^2 \int_k\!\frac{\partial}{\partial k^\mu}{\cal F}^{(3)}_\mu(k);&
	{\cal F}^{(3)}_\mu(k) &=  k_\mu\Delta(k^2)Y(k^2).
\label{F3QCD}
\end{align}

In summary, the above demonstration establishes that, under the pivotal assumption of the absence of $q^2$-type of poles in the 
form factors of the fundamental vertices,  
\begin{align}
\widetilde{\Pi}^{(i)}(0)&= 0;\quad i=1,2,3& &\Longrightarrow& \widetilde{\Pi}(0) &= \sum_{i=1}^{3}\widetilde{\Pi}^{(i)}(0) =0,
\label{all0}	
\end{align}
which, since $\widetilde\Delta^{-1}(q^2) = q^2 + i \widetilde{\Pi}(q^2)$, leads to the conclusion that  
\be 
\widetilde{\Delta}^{-1}(0) =0. 
\label{hatD0}
\ee

In order to extract from \1eq{hatD0} the behavior of the 
conventional (quantum) quantity ${\Delta}^{-1}(q^2)$ at the origin, one additional step is necessary.
Specifically, we must employ the crucial identity \1eq{propBQI}, which, at $q=0$ yields
\be 
\Delta^{-1}(0) = \frac{\widetilde{\Delta}^{-1}(0)}{1+G(0)}.
\label{D1G}
\ee
If we now introduce the additional assumption that $1+G(0)$ is finite for every $\xi$ (see discussion below), 
then we reach the final conclusion that, in the absence of poles,  
\be 
\Delta^{-1}(0) = 0.
\ee

\subsection{\label{subsec:ar}Additional remarks}  

\renewcommand{\theenumi}{\arabic{enumi}}
\begin{enumerate}
	\item It is important to recognize that  the proofs elaborated in the previous subsections are valid for {\it any value of the gauge-fixing parameter $\xi$} within the class of the linear covariant gauges ($R_{\xi}$); indeed, at no moment has it been necessary to choose a specific value for $\xi$.
	\item Related to this point, in a general $R_\xi$ gauge there exists a relation between $G$ and the ghost dressing function $F$ which reads~\cite{Binosi:2013cea}
		\begin{align}
			F^{-1}(q^2)=1+G(q^2)+L(q^2)+\xi K(q^2),
			\label{funrelxi}
		\end{align}	
where $L$ is the $q_{\mu}q_{\nu}/q^2$ component of the function mentioned right after \1eq{propBQI}, 
and $K$ originates from the coupling of the antighost to a certain anti-BRST source, 
necessary for formulating the theory in the background field method (or, equivalently, rendering the conventional theory anti-BRST invariant)~\cite{Binosi:2013cea}. 
When $\xi=0$,  we know that $F(0)\neq 0$ while $L(0)=0$~\cite{Aguilar:2009pp}; therefore, from \1eq{funrelxi} one obtains  
		\begin{align}
			F^{-1}(0)=1+G(0),
			\label{funrel}
		\end{align}
	which ensures that the quantity $1+G(0)$ is finite in the Landau gauge.
For $\xi\neq0$ the situation is more complicated. Recent analytical studies~\cite{Aguilar:2015nqa,Huber:2015ria} indicate that in this case the ghost dressing function goes to zero, 
which, in turn, would suggest that the l.h.s. of~\1eq{funrelxi} diverges. 
At the same time, however, they also show that the gluon propagator continues to saturate in the IR, a result that has been confirmed by lattice simulations~\cite{Bicudo:2015rma}. If the behavior of the ghost dressing function is to be confirmed, these results would point towards some highly nontrivial dynamics in the ghost sector, which would ensure that the function $K$ (and possibly $L$) appearing in~\1eq{funrelxi} will cancel the divergence of the ghost dressing function, leaving a finite $1+G(0)$ when $\xi\neq 0$.
	\item  Note that in the demonstrations presented  no particular form for the propagators $\Delta$ (or $D$) appearing in the one- and two-loop dressed diagrams has been used. In fact, even if one were to {\it assume} that the gluon propagator circulating in them is of a massive type\footnote{For instance, one could consider a propagator of the Cornwall type~\cite{Cornwall:1981zr}, $\Delta(q^2) = 1/[q^2+m^2(q^2)]$, or of the Stingl form~\cite{Stingl:1985hx}, \mbox{$\Delta(q^2)=c(1+aq^2)/[(q^2+m^2)^2+b^2]$}, with $a,\ b,\ c$ suitable parameters.}, the seagull identity would still annihilate the individual contributions $\widetilde{\Pi}^{(1,2,3)}(0)$, yielding $\Delta^{-1}(0)=0$, in contradiction with the original assumption. The main lesson drawn from this observation is that the theory, when properly treated, resists the generation of a `mass', due to the operation of a very subtle cancellation mechanism. Of course, the realization of this mechanism hinges crucially on the absence of poles in the vertices of the theory; it is precisely the relaxation of this assumption that will eventually permit the emergence of infrared finite gluon propagators, as explained in the continuation of this article.
	\item We emphasize that no specific Ansatz for any of the fully-dressed vertices appearing in the above derivations has been employed. This is to be contrasted with the original demonstration of the seagull cancellation presented in~\cite{Aguilar:2009ke}: there, a gauge-technique-inspired Ansatz  was used for $\widetilde{\Gamma}_{\mu\alpha\beta}(q,r,p)$, satisfying (by construction) the first Abelian STI of \1eq{BQids}, which, when inserted into graph ($a_1$),  activated the original form of the seagull identity, \1eq{seaold}.
This is particularly relevant in the 
case of the four-gluon vertex, where, due to the complexity of the corresponding STI, 
the construction of such an Ansatz has never been presented in the literature.
	\item It is instructive to demonstrate the vanishing of $\widetilde{\Pi}^{(1)}(0)$ and $\widetilde{\Pi}^{(2)}(0)$ using the form of the relevant form factors, \1eq{Awi} and \1eq{theA}, to activate \1eq{seaold}. In the case of $\widetilde{\Pi}^{(2)}(0)$, we substitute directly \1eq{ghq0} and \1eq{Awi} into $a_3(0)$  ($r \to k$)
		\begin{align}
			\widetilde a_3(0) = 2ig^2 C_A\int_k\! k^2 D^2(k^2) \frac{\partial}{\partial k^2}D^{-1}(k^2) = -2i g^2 C_A\int_k\! k^2 \frac{\partial}{\partial k^2} D(k^2),
		\end{align}
	which, when added to $a_4(0)$, triggers~\1eq{seaold} with $f(k^2) = D(k^2)$. The case of $\widetilde{\Pi}^{(1)}(0)$ proceeds in a similar fashion, but is operationally slightly more involved. Substituting \2eqs{Gq0}{theA} into $a_1(0)$, it is straightforward to establish that the contribution from $\widetilde{A}_{10}$ vanishes, whereas $\widetilde{A}_{6}$, $\widetilde{A}_{13}$, and $\widetilde{A}_{14}$ combine to yield
		\be
			\widetilde a_1(0) = -g^2C_A (d-1) \left[ 2 \int_k k^2 \frac{\partial}{\partial k^2}\Delta(k^2) + \int_k\Delta(k^2) \right].
		\ee
		Then, using that 
		\be
			\widetilde a_2(0) = -g^2C_A (d-1)^2  \int_k\Delta(k^2), 
		\ee
		we find that 
		\be
		d\widetilde\Pi^{(1)}(0) = -2 g^2C_A (d-1) \left[ \int_k k^2 \frac{\partial}{\partial k^2} \Delta(k^2) + \frac{d}{2} \int_k\Delta(k^2)\right],
		\ee
		which triggers \1eq{seaold}, with \mbox{$f(k^2)=\Delta(k^2)$}.
		 		
	Note, however, that the application of the above procedure at the two-loop dressed level, in order to demonstrate the vanishing of $\widetilde{\Pi}^{(3)}(0)$, would be completely impractical, given that, as mentioned at the end of subsection~\ref{subsec:wiavff}, expressions analogous to \1eq{Awi} and \1eq{theA} for the relevant form factors of $\widetilde{\Gamma}^{mnrs}_{\mu\alpha\beta\gamma}$ are rather difficult to derive. Instead, the use of the compact version of the seagull identity, \1eq{seagull},  requires only the global form of the corresponding WI that $\widetilde{\Gamma}^{mnrs}_{\mu\alpha\beta\gamma}$ satisfies, making no reference whatsoever to its tensorial decomposition. This fact, in turn, exemplifies the advantages of the present formulation, and allows one to explore important aspects of the two-loop dressed structure, which otherwise would have been unattainable.    
\end{enumerate}

\section{\label{sec:r}Renormalization} 
In the previous section,   
the behavior of the gluon propagator at the origin has been 
derived using bare (unrenormalized) quantities. It is therefore important to 
establish that the main conclusion, namely the vanishing of $\Delta^{-1}(0)$
in the absence of massless poles, persists after renormalization, \ie that $\Delta^{-1}_{\s R}(0)=0$.  

Let us start by stating the renormalization conditions in the quantum sector of the theory:
\begin{align}
	\Delta_{\s R} &=  Z^{-1}_{\s Q} \Delta(q^2);&
	D_{\s R}&= Z^{-1}_{c} D(q^2);&   
	g_{\!\s R} &= Z_g^{-1} g,
	\label{renprop}\\
	{\Gamma}_{\!\s R}^{\mu} &= {Z}_1 {\Gamma}^{\mu};&\qquad  
{\Gamma}_{\!\s R}^{\mu\alpha\beta} &= {Z}_3 {\Gamma}^{\mu\alpha\beta};&
{\Gamma}_{\!{\s R}\,\mu\alpha\beta\nu}^{mnrs} &= {Z}_4 {\Gamma}_{\mu\alpha\beta\nu}^{mnrs}.
	\label{renconst}
\end{align}
Note in particular that $Z^{-1}_{\s Q}$ is reserved for the renormalization of the quantum gauge field $Q$, 
whereas the corresponding constant renormalizing the background field $B$, to be introduced below, will be denoted by $Z^{-1}_{\s B}$. 

In the quantum sector, the constraints relating the above constants 
is exactly the same as in the conventional covariant gauges~\cite{Pascual:1984zb}; specifically, 
the standard STIs of the theory enforce the validity of    
\be
Z_g = Z_1 Z_{\s Q}^{-1/2} Z_c^{-1} = Z_3  Z_{\s Q}^{-3/2} = Z_4^{1/2} Z_{\s Q}^{-1}.
\label{STIrel}
\ee

Consider now the background and mixed quantum-background sectors. The relevant two-point functions are renormalized as 
\begin{align}
	\widehat\Delta_{\s R} &= Z^{-1}_{\s B} \widehat\Delta;&
	\widetilde\Delta_{\s R} &= {\cal Z}^{-1} \widetilde\Delta;&
	G_{\s R}&=Z_\s{G}G,
\end{align}
which,  due to the residual background symmetry~\cite{Abbott:1980hw,Abbott:1981ke}, 
or relations such as \1eq{propBQI} and \1eq{funrelxi}, satisfy   
\begin{align}
	Z_g &= Z^{-1/2}_{\s B};&
	{\cal Z}&=Z^{1/2}_{\s Q} Z^{1/2}_{\s B};&
	Z_{\s G} &= Z^{-1/2}_{\s Q} Z^{1/2}_{\s B};&
	Z_{\s G} &= Z_c Z_1^{-1}={\cal Z}Z_\s{Q}^{-1}.
\end{align} 

Similarly, the renormalization constants of the three vertices involving one background gluon (in the $q$-channel) are defined as 
\begin{align}
	\widetilde{\Gamma}_{\!\s R}^{\mu} &= \widetilde{Z}_1 \widetilde{\Gamma}^{\mu};&\widetilde{\Gamma}_{\!\s R}^{\mu\alpha\beta} &=\widetilde{Z}_3 \widetilde{\Gamma}^{\mu\alpha\beta};&
\widetilde{\Gamma}_{\!{\s R}\,\mu\alpha\beta\nu}^{mnrs} &= \widetilde{Z}_4 \widetilde{\Gamma}_{\mu\alpha\beta\nu}^{mnrs},
\label{renconst2}
\end{align}
and the corresponding Abelian STIs impose the crucial conditions,  
\be
\widetilde{Z}_1 = Z_{c}; \qquad 
\widetilde{Z}_3 = Z_{\s Q}; \qquad 
\widetilde{Z}_4 = Z_3, 
\label{rencond}
\ee
which must be preserved by the renormalization procedure, and in particular by the 
renormalization scheme chosen. 
We implicitly assume that all pertinent renormalization conditions are imposed at a renormalization point that lies in a region 
where the form of the relevant Green's functions, and especially of the gluon propagator, are under control,  
\ie where perturbative considerations are still applicable.

Given the relations above, let us see how the SDE for $\widetilde\Delta$ is renormalized. One has
\bea
d\widetilde{\Pi}^{(1)}(q^2) &=& \widetilde a_1(q^2) + \widetilde a_2(q^2) 
\nonumber\\
&=& g^2C_A \left[
\frac{1}{2}\int_k \Gamma_{\mu\alpha\beta}^{(0)}\Delta^{\alpha\rho}(k)
\Delta^{\beta\sigma}(k+q)\widetilde{\Gamma}^\mu_{\sigma\rho}
+ (d-1) \int_k\Delta^{\alpha}_{\alpha}(k) \right]
\nonumber\\
&=&
Z_g^2 Z_{\s Q} g^2_{\!\s R} C_A  \left[
\frac{1}{2}\int_k \Gamma_{\mu\alpha\beta}^{(0)}\Delta^{\alpha\rho}_{\!\s R}(k)\Delta^{\beta\sigma}_{\!\s R}(k+q)
\widetilde{\Gamma}^\mu_{{\!\s R}\,\sigma\rho}
+ (d-1) \int_k\Delta^{\alpha}_{{\!\s R}\,\alpha}(k) \right]
\nonumber\\
&=& Z_g^2 Z_{\s Q} \left[ a_1^{\!\s R}(q^2) + a_2^{\!\s R}(q^2) \right]
\nonumber\\
&=& Z_g^2 Z_{\s Q} \,d\,\widetilde{\Pi}^{(1)}_{\!\s R}(q^2),
\eea
and, similarly,
\begin{align}
	d \widetilde{\Pi}^{(2)}(q^2) &=
a_3(q^2) + a_4(q^2) = Z_g^2 Z_{c} \left[ a_3^{\!\s R}(q^2) + a_4^{\!\s R}(q^2) \right] = Z_g^2 Z_{c} \, d \widetilde{\Pi}^{(2)}_{\!\s R}(q^2)
\nonumber\\
d \widetilde{\Pi}^{(3)}(q^2) &=
a_5(q^2) + a_6(q^2) = Z_g^4 Z_3^{-1} Z^3_{\s Q} \left[ a_5^{\!\s R}(q^2) + a_6^{\!\s R}(q^2) \right] = 
Z_g^4 Z_3^{-1} Z^3_{\s Q} \,d \widetilde{\Pi}^{(3)}_{\!\s R}(q^2)
\end{align}
where we have introduced the combinations 
\begin{align}
Z_3&={\cal Z} Z_g^2 Z_{\s Q};&  Z_1&={\cal Z} Z_g^2 Z_{c};&  
Z_4&={\cal Z} Z_g^4 Z_3^{-1} Z^3_{\s Q}.
\end{align}

Combining all the above equations, we find the relation
\be
\widetilde\Delta^{-1}_{\!\s R}(q^2) = {\cal Z}q^2 + i \left[ Z_3 \widetilde{\Pi}^{(1)}_{\!\s R}(q^2) +  
Z_1 \widetilde{\Pi}^{(2)}_{\!\s R}(q^2) + Z_4 \widetilde{\Pi}^{(3)}_{\!\s R}(q^2)\right],
\ee
which, under the assumption of a finite $1+G_{\s R}(0)$, implies that $\Delta^{-1}_{\s R}(0) = 0$.

Let us go one step further and
impose the momentum subtraction (MOM) renormalization condition $\widetilde\Delta^{-1}_{\s R}(\mu^2) = \mu^2$, so that 
\be
{\cal Z} = 1 - \frac{i}{\mu^2} \left[Z_3 \widetilde{\Pi}^{(1)}_{\!\s R}(\mu^2) + Z_1 \widetilde{\Pi}^{(2)}_{\!\s R}(\mu^2)
+ Z_4 \widetilde{\Pi}^{(3)}_{\!\s R}(\mu^2)\right].
\ee
At this point one has
\begin{align}
\widetilde\Delta^{-1}_{\s R} (q^2) &= q^2 +  i \left[Z_3 \widetilde{\Pi}^{(1)}_{\!\s R}(q^2) +  
Z_1 \widetilde{\Pi}^{(2)}_{\!\s R}(q^2) + Z_4 \widetilde{\Pi}^{(3)}_{\!\s R}(q^2)\right]
\nonumber\\
&-  \frac{q^2}{\mu^2}
i\left[Z_3 \widetilde{\Pi}^{(1)}_{\!\s R}(\mu^2) + Z_1 \widetilde{\Pi}^{(2)}_{\!\s R}(\mu^2)
+ Z_4 \widetilde{\Pi}^{(3)}_{\!\s R}(\mu^2)\right],
\end{align}
so that, once again, $\Delta^{-1}_{\s R}(0) = 0$.


\section{\label{sec:etscvwmp} Evading the seagull cancellations: vertices with poles} 

In order to obtain an infrared finite gluon propagator self-consistently, one needs 
to introduce poles in the vertices, and, in particular, in the 
channel that is associated with the momentum flowing into the gluon SDE, where 
these vertices are eventually inserted. To be precise, 
and following the conventions of Figs.~\ref{fig:a1-a2},~\ref{fig:a3-a4}, and~\ref{fig:a5-a6}, 
(some of) the vertices $\widetilde{\Gamma}_{\mu\alpha\beta}$, $\widetilde{\Gamma}_{\mu}$, and 
$\widetilde{\Gamma}_{\mu\alpha\beta\gamma}^{mnrs}$  must contain pole terms of the form $q^{\mu}/q^2$.
The purpose of this section is to study how the 
inclusion of such terms circumvents the seagull identity, thus allowing for the possibility of $\Delta^{-1}(0) \neq 0$.

Specifically, let us assume that some of the form factors now contain two distinct parts, which will de indicated 
by a superscript `${\bf p}$' (for `pole' parts) or `${\bf np}$' (for `no-pole' parts). In that sense, whereas before none of the form factors contained poles, now we can say that, using the same notation introduced in~\2eqs{ghgen}{glgen}, 
\begin{align}
\widetilde {\cal A}_1  &= \widetilde {\cal A}_1^{\bf np} + \widetilde {\cal A}_1^{\bf p};& \widetilde {\cal A}_2  &= \widetilde {\cal A}_2^{\bf np},
\nonumber\\
\widetilde {A}_i  &= \widetilde {A}_i^{\bf np} + \widetilde {A}_i^{\bf p},\quad i=1,...,5 ;& \widetilde {A}_i  &= \widetilde {A}_i^{\bf np},\quad i=6,...,14.& 
\label{Aipnp}
\end{align}
The fact that only the longitudinally-coupled form factors are allowed to contain  pole parts is dictated by the physical requirement that they should act like `dynamical Nambu-Goldstone bosons', and decouple from physical observables~\cite{Jackiw:1973tr,Jackiw:1973ha,Cornwall:1973ts,Eichten:1974et,Poggio:1974qs,Aguilar:2011xe,Ibanez:2012zk}.

\begin{figure}[!t]
\includegraphics[scale=0.65]{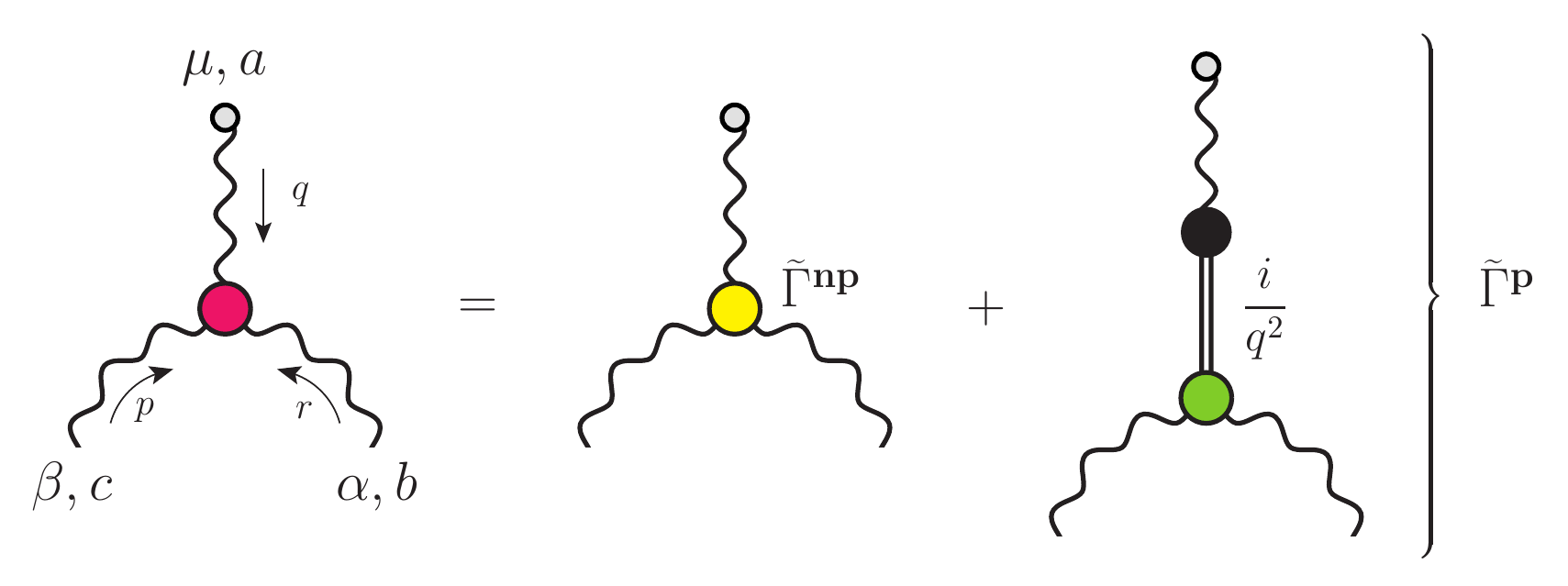}
\caption{\label{fig:p-np}(Color online) The no-pole and pole parts of the $BQ^2$ gluon vertex. Analogous decompositions hold for the $Bc\bar c$ and the $BQ^3$ vertices.}
\end{figure}

Thus, one may cast the nonperturbative vertices in the form~(see \fig{fig:p-np})
\begin{align}
\widetilde{\Gamma}_{\mu\alpha\beta}(q,r,p) &= \g_{\mu\alpha\beta}(q,r,p) + \gp_{\mu\alpha\beta}(q,r,p),
\nonumber\\
\widetilde{\Gamma}_{\mu}(q,r,p) &= \g_{\mu}(q,r,p) + \gp_{\mu}(q,r,p),
\nonumber\\
\widetilde{\Gamma}_{\mu\alpha\beta\gamma}^{mnrs}(q,r,p,t) &= \widetilde{\Gamma}_{\mu\alpha\beta\gamma}^{{\bf np},mnrs}(q,r,p,t) + 
\widetilde{\Gamma}_{\mu\alpha\beta\gamma}^{{\bf p},mnrs}(q,r,p,t),
\label{altwr}
\end{align} 
where, due to the condition of longitudinality, one can write in full generality
\begin{align}
{\gp}_{\mu\alpha\beta}(q,r,p) &= \frac{q_\mu}{q^2}{\widetilde C}_{\alpha\beta}(q,r,p),& 
\nonumber\\	
{\gp}_{\mu}(q,r,p) & = \frac{q_\mu}{q^2}{\widetilde C}(q,r,p), 
\nonumber\\
\widetilde{\Gamma}_{\mu\alpha\beta\gamma}^{{\bf p},mnrs}(q,r,p,t) & = \frac{q_\mu}{q^2}{\widetilde C}^{mnrs}_{\alpha\beta\gamma}(q,r,p,t).
\label{UIB}
\end{align}

The precise way how the form factors of \1eq{Aipnp} comprise $\g$ and $\gp$, and in particular the 
functions  ${\widetilde C}$, may be easily worked out. For example, in the 
case of $\widetilde{\Gamma}_{\mu}$, which is the vertex with the  simplest tensorial structure, we simply obtain  
\begin{align}
\g_{\mu}(q,r,p) &= \widetilde {\cal A}_1^{\bf np} q_{\mu} + \widetilde {\cal A}_2^{\bf np} r_{\mu},&
\nonumber\\
\gp_{\mu}(q,r,p) & = \widetilde {\cal A}_1^{\bf p} q_{\mu}\qquad \Longrightarrow \qquad\widetilde {\cal A}_1^{\bf p}\equiv \frac{{\widetilde C}(q,r,p)}{q^2}
\label{compr}
\end{align}

Of course, in order to keep the BRST invariance intact, we demand that all STIs 
maintain their exact form in the presence of these poles; therefore, \1eq{BQids} will now read  
\begin{align}
	q^\mu \g_{\mu\alpha\beta}(q,r,p) + \widetilde{C}_{\alpha\beta}(q,r,p) &= i\Delta_{\alpha\beta}^{-1}(r) - i\Delta_{\alpha\beta}^{-1}(p),
\nonumber\\
q^\mu\g_{\mu}(q,r,p) +\widetilde{C}(q,r,p) &= iD^{-1}(r^2) - iD^{-1}(p^2), 
\nonumber\\
	q^\mu  \widetilde{\Gamma}_{\mu\alpha\beta\gamma}^{{\bf np},mnrs}(q,r,p,t) + \widetilde{C}_{\alpha\beta\gamma}^{mnrs}(q,r,p,t)
&= f^{mse}f^{ern} \Gamma_{\alpha\beta\gamma}(r,p,q+t) + f^{mne}f^{esr}\Gamma_{\beta\gamma\alpha}(p,t,q+r) 
\nonumber\\
	&+ f^{mre}f^{ens} \Gamma_{\gamma\alpha\beta}(t,r,q+p).
	\label{BQinew}
\end{align}

Note that if $\widetilde{\Gamma}_{\mu\alpha\beta}(q,r,p)$ contains poles in $q^2$, by virtue of the 
corresponding `background-quantum identity'~\cite{Binosi:2008qk} so does  
the conventional ($Q^3$) three-gluon vertex $\Gamma_{\alpha\beta\gamma}(q,r,p)$. However, 
the r.h.s. of the third STI in \1eq{BQinew} contains no such poles,   
because $q$ never appears in the arguments of the $\Gamma$s alone, but rather in the combinations $q+t$, $q+r$, or $q+p$.  

At this point it should be clear that  $\g$ represents precisely
the part of  the total vertex $\widetilde{\Gamma}$ that  will enter in
the calculation  of $\widetilde  \Pi(0) g_{\mu\nu}$,  and consequently
will participate in  the seagull cancellation. On the  other hand, the
term  with the  massless pole  in $q^2$  will contribute  to the  term
$\widetilde \Pi(0)  q_{\mu}q_{\nu}/q^2$, which is not  involved in the
seagull cancellation. Of course, since the original STIs of \1eq{BQids}, 
now replaced by the equivalent set given in \1eq{BQinew}, remain intact, the block-wise
transversality property  of~\1eq{blockwise} is  automatically incorporated in the final answer.
Therefore, the total  contribution of each block to the $g_{\mu\nu}$  part (after  the seagull
cancellation) will  be exactly  equal (and opposite  in sign)  to that
proportional to $q_{\mu}q_{\nu}/q^2$.

In order to appreciate the above points in detail, 
let us study the $q=0$ limit of the gluon self-energy as done previously in Sec.~\ref{sec:awiotpv}. 
To this end we need to derive the equivalent of~\3eqs{BQ2-id}{Bc2-id}{BQ3-id}. This can be done by taking directly a Taylor expansion around $q=0$ of both sides of Eqs.\noeq{BQinew}, as they are both regular in this limit. Evidently, the zeroth order term vanishes in all three cases, 
\begin{align}
	&\widetilde{C}_{\alpha\beta}(0,r,-r) =0;&
	&\widetilde{C}(0,r,-r) =0;&
	&\widetilde{C}_{\alpha\beta\gamma}^{mnrs}(0,r,p,-p-r) = 0,
	\label{C0s}
\end{align}
whereas the first order terms yield for the $BQ^2$ vertex
\begin{align}
	\g_{\mu\alpha\beta}(0,r,-r) &= -i\frac{\partial}{\partial r^\mu}\Delta^{-1}_{\alpha\beta}(r) - \left\lbrace\frac{\partial}{\partial q^\mu}\widetilde{C}_{\alpha\beta}(q,r,-r-q)\right\rbrace_{q=0};\nonumber \\
	\g_{\mu\alpha\beta}(0,-p,p) &= i\frac{\partial}{\partial p^\mu}\Delta^{-1}_{\alpha\beta}(p) - \left\lbrace\frac{\partial}{\partial q^\mu}\widetilde{C}_{\alpha\beta}(q,-p-q,p)\right\rbrace_{q=0},
	\label{IAbelian STIBQ2pole}
\end{align}
for the $B\bar c c$ vertex
\begin{align}
	\g_\mu(0,r,-r)  &= -i\frac{\partial}{\partial r^\mu} D^{-1}(r^2)- \left\lbrace\frac{\partial}{\partial q^\mu}\widetilde{C}(q,r,-r-q)\right\rbrace_{q=0};\nonumber \\ 
	\g_\mu(0,-p,p)  &= i\frac{\partial}{\partial p^\mu} D^{-1}(p^2)- \left\lbrace\frac{\partial}{\partial q^\mu}\widetilde{C}(q,-p-q,p)\right\rbrace_{q=0} ,
\label{WIsnew}
\end{align}
and, finally, for the $BQ^3$ vertex
\begin{align}
	\widetilde{\Gamma}^{{\bf np},mnrs}_{\mu\alpha\beta\gamma}(0,r,p,-r-p) & = \left(f^{mne}f^{esr}\frac{\partial}{\partial r^\mu} + f^{mre}f^{ens}\frac{\partial}{\partial p^\mu}\right)\Gamma_{\alpha\beta\gamma}(r,p,-r-p)\nonumber\\
	& - \bigg\lbrace \frac{\partial}{\partial q^\mu}\widetilde{C}_{\alpha\beta\gamma}^{mnrs}(q,r,p,-q-r-p)\bigg\rbrace_{q=0}, \nonumber \\
	\widetilde{\Gamma}^{{\bf np},mnrs}_{\mu\alpha\beta\gamma}(0,-r,-p,r+p) & = -\left(f^{mne}f^{esr}\frac{\partial}{\partial r^\mu} + f^{mre}f^{ens}\frac{\partial}{\partial p^\mu}\right)\Gamma_{\alpha\beta\gamma}(-r,-p,r+p)\nonumber\\
	& + \bigg\lbrace \frac{\partial}{\partial q^\mu}\widetilde{C}_{\alpha\beta\gamma}^{mnrs}(-q,-r,-p,q+r+p)\bigg\rbrace_{q=0}.
	\label{WIsnew2}
\end{align}
These identities, in turn, provide symmetry constraints for the $\widetilde C$ functions. In particular, 
in the $BQ^2$ and $B\bar c c$ cases, the $\widetilde C$ is invariant upon inversion of all momenta, 
whereas it is antisymmetric when inverting the last two momenta (and the corresponding indices in the $BQ^2$ case). 
For the $BQ^3$ vertex instead, $\widetilde C$ behaves with respect to its arguments as a conventional three-gluon vertex. 

Let us now repeat the calculation of Sec.~\ref{sec:tgseato} following the exact same logic; specifically, the vertex contributions that survive the limit $q=0$ in the various self-energy diagrams must be replaced by the l.h.s. of the corresponding WIs. However, the crucial difference now is that the WIs to employ are those given in \3eqs{IAbelian STIBQ2pole}{WIsnew}{WIsnew2}, which, in addition to the terms already present in \3eqs{BQ2-id}{Bc2-id}{BQ3-id}, contain the derivatives of the functions $\widetilde{C}$.      As a result, whereas the first type of terms will trigger again the seagull identities and vanish exactly as before,  the contributions originating from  $\widetilde{C}$ will escape the total annihilation. In particular, it is fairly straightforward to show that    
\begin{align}
	d\widetilde{\Pi}^{(1)}(0) &= -\frac{1}{2}g^2C_A \int_k \Gamma_{\mu\alpha\beta}^{(0)}(0,k,-k)\Delta^{\alpha\rho}(k)\Delta^{\beta\sigma}(k)
\left\{\frac{\partial}{\partial q^\mu} \widetilde{C}_{\sigma\rho}(q,-k-q,k)\right\}_{q=0}, \label{0a1a2mass} \\
d\widetilde{\Pi}^{(2)}(0) &= -g^2C_A \int_k \Gamma_{\mu}^{(0)}(0,k,-k) D^{2}(k^{2})\left\{\frac{\partial}{\partial q^\mu}\widetilde{C}(q,-k-q,k)\right\}_{q=0}, 
\label{0a3a4mass}
\end{align} 
and 
\begin{align}
d\widetilde{\Pi}^{(3)}(0)\delta^{ab} &= \frac{1}{6}g^4\Gamma_{\mu\alpha\beta\gamma}^{(0)amnr}\int_k\int_\ell \Delta^{\alpha\rho}(k+\ell)\Delta^{\beta\sigma}(\ell)\Delta^{\gamma\tau}(k)\nonumber \\
&\times
\left\{ \frac{\partial}{\partial q^\mu}\widetilde{C}_{\tau\sigma\rho}^{brnm}(q,-k-q,-\ell,k+\ell)\right\}_{q=0} \nonumber \\
&+ i{\cal N}_{\mu\alpha\beta\gamma}\delta^{ab} \int_k Y_\delta^{\alpha\beta}(k)\Delta^{\gamma\tau}(k)
\Delta^{\lambda\delta}(k)\left\lbrace\frac{\partial}{\partial q^\mu} \widetilde{C}_{\tau\lambda}(q,-k-q,k)\right\rbrace_{q=0}.
\label{0a5a6mass}
\end{align}
Thus, the presence of $1/q^2$ poles allows for a non-vanishing value of the self-energy at $q=0$, therefore providing the possibility for $\widetilde{\Delta}^{-1}(0)\neq0$. 

Let us end this section by proving that one recovers the same answer for $\widetilde{\Pi}(0)$ by considering directly the part of the self-energy proportional to $q_{\mu}q_{\nu}/q^2$, where, of course, no seagull identity is operating. To work out a concrete example in detail, consider the one-loop dressed gluon diagrams; in this case, the desired contribution is obtained by simply replacing inside graph $(\widetilde a_1)$ the full vertex by its pole part. 
One obtains 
\be
{\frac{q_\mu q_\nu}{q^2}} \widetilde{\Pi}^{(1)}(q^2) =\frac{1}{2}g^2C_A \frac{q_\nu}{q^2}\int_k \Gamma_{\mu\alpha\beta}^{(0)}(q,k,-k-q)\Delta^{\alpha\rho}(k)\Delta^{\beta\sigma}(k+q)\widetilde{C}_{\sigma\rho}(q,-k-q,k)+\cdots,
\ee
where the dots indicates terms that vanish as $q\to0$; then one immediately has 
\be
\widetilde{\Pi}^{(1)}(q^2) = \frac{1}{2}g^2C_A \frac{q^\mu}{q^2} \int_k \Gamma_{\mu\alpha\beta}^{(0)}(q,k,-k-q)\Delta^{\alpha\rho}(k)\Delta^{\beta\sigma}(k+q)\widetilde{C}_{\sigma\rho}(q,-k-q,k)+\cdots.
\ee
We can now expand $\widetilde{C}$ around $q=0$ and use the fact that  $\widetilde{C}_{\sigma\rho}(0,-k,k)=0$ [see~\1eq{C0s}], to obtain
\begin{align}
	\widetilde{\Pi}^{(1)}(q^2) &= \frac{1}{2}g^2C_A \frac{q^\mu q^\lambda}{q^2} X_{\mu\lambda},
	\label{X}
\end{align}
with
\begin{align}
	X_{\mu\lambda} &= \int_k \Gamma_{\mu\alpha\beta}^{(0)}(0,k,-k)\Delta^{\alpha\rho}(k)\Delta^{\beta\sigma}(k) \left\{\frac{\partial}{\partial q^\lambda} \widetilde{C}_{\sigma\rho}(q,-k-q,k)\right\}_{q=0}.
	\end{align}
Then, since $X_{\mu\lambda}$ has no dependence on $q$, it can only be proportional to the metric tensor $g_{\mu\lambda}$, namely one has
\be
dX_{\mu\lambda} = g_{\mu\lambda} \int_k \Gamma_{\nu\alpha\beta}^{(0)}(0,k,-k)\Delta^{\alpha\rho}(k)\Delta^{\beta\sigma}(k) 
\bigg\lbrace\frac{\partial}{\partial q_\nu}\widetilde{C}_{\sigma\rho}(q,-k-q,k)\bigg\rbrace_{q=0},
\ee 
which, upon substitution into~\1eq{X}, gives the announced equality with~\1eq{0a1a2mass}. An exactly analogous procedure may be followed for the remaining contributions, thus establishing explicitly the transversality of the gluon self-energy even in the presence of massless poles.

\section{\label{sec:num} Further considerations and Numerical analysis}

The WIs given in \3eqs{IAbelian STIBQ2pole}{WIsnew}{WIsnew2} furnish certain interesting relations among the form factors of the vertices that satisfy them. In order to derive them, let us consider a general scalar function $f(q,r,p)$, which is antisymmetric under  $r\leftrightarrow p$, and expand it around $q=0$ (and $p=-r$). Since in that case $f(0, r,-r) =0$, we have that  
\be
f(q,r,p)=2(q\!\cdot\!r)  f^{\prime}(r,-r)  +{\cal O}(q^2),
\label{Taylor}
\ee
where the prime denotes differentiation with respect to $(r+q)^2$ and subsequently taking the 
limit $q\to 0$, \ie 
\be
f^{\prime}(r,-r) \equiv  \lim_{q\to 0} \frac{\partial}{\partial\, (r+q)^2}f(q,r,-r-q) .
\label{Der}
\ee
Evidently, due to Lorenz invariance, $f^{\prime}(r,-r)=f'(r^2)$.

Then, if we expand \1eq{WIsnew} around $q=0$, the analogue of \1eq{Awi} may be obtained after using that 
\be
\left\lbrace\frac{\partial}{\partial q^\mu}\widetilde{C}(q,r,p)\right\rbrace_{q=0} = 2 r_{\mu} \widetilde{C}^{\prime}(r^2),
\label{Der2}
\ee
and so, 
\be
\widetilde{\cal A}^{\bf np}_2(r^2) = -2 \left[ i\frac{\partial}{\partial r^2}D^{-1}(r^2) + \widetilde{C}^{\prime}(r^2)\right].
\label{Awinp}
\ee
The above argument may be extended directly to the case of the three-gluon vertex; to simplify the situation,   
we assume that out of the five possible tensorial structures of $\widetilde{C}_{\sigma\rho}$ 
only the one proportional to $g_{\sigma\rho}$ develops a pole in $q^2$; at the level of 
$\gp_{\mu\alpha\beta}$ this is equivalent to the statement that only ${A}_1^{\bf p}\equiv \widetilde{C}_1/{q^2} \neq 0$.
Then,
\be
\left\lbrace\frac{\partial}{\partial q^\mu}\widetilde{C}_{\alpha\beta}(q,r,p)\right\rbrace_{q=0} = 
2 r_{\mu} g_{\alpha\beta}\widetilde{C}^{\prime}_1(r^2), 
\ee
and the first identity in \1eq{theA} assumes the form   
\be
\widetilde A_6^{\bf np}(r^2) = 2 \left[\frac{\partial}{\partial r^2}\Delta^{-1}(r^2) - \widetilde{C}_1^{\prime}(r^2)\right].
\label{A6np}
\ee

In order to gain a basic quantitative understanding of some of the relations derived here, we next carry out a numerical analysis, under a number of simplifying assumptions. In particular, we assume that the strength of the poles coming from ${\gp}_{\mu}(q,r,p)$ and $\widetilde{\Gamma}_{\mu\alpha\beta\gamma}^{{\bf p},mnrs}(q,r,p,t)$ is suppressed with respect to that of ${\gp}_{\mu\alpha\beta}(q,r,p)$, and may therefore be neglected. In addition, we will keep as before only the component $\widetilde{C}_1$, so that \1eq{0a1a2mass} yields
\begin{align}
	d\widetilde{\Pi}^{(1)}(0) &=2g^2C_A\int_k\!\left[(d-1)k^2\Delta^2(k^2)\right]\widetilde{C}'_1(k^2).
\end{align}
Considering for simplicity the Landau gauge, $\xi=0$, one finds (Euclidean space)
\begin{align}
	\widetilde\Delta^{-1}(0) = -2 g^2 C_A \frac{d-1}{d} \int_k k^2 \Delta^2(k^2) \widetilde{C}_1^{\prime}(k^2),
\label{num4}
\end{align}
or using~\2eqs{D1G}{funrel}, introducing spherical coordinates, and setting $k^2=y$, $d=4$, and $\alpha_s = g^2/4\pi$,
\be
\Delta^{-1}(0) = -\frac{3 C_A \alpha_s}{8\pi} F(0)\int_0^{\infty}\!\mathrm{d}y\, y^2 \Delta^2(y) \widetilde{C}_1^{\prime}(y),
\label{num5}
\ee
Evidently, given that $\Delta^{-1}(0)$ and $F(0)$ are positive quantities, the function 
$\widetilde{C}_1^{\prime}(y)$ must be such that, when inserted in \1eq{num5}, it will compensate for the overall minus sign. 
One way to accomplish this is by assuming that $\widetilde{C}_1^{\prime}(y)$ is negative throughout the entire range of momenta.
Alternatively, one may envisage a type of function that changes its sign, 
having positive and negative supports that are appropriately distributed 
with respect to the function $y^2 \Delta^2(y)$, eventually furnishing more negative 
than positive contribution to the integral of \1eq{num5}. To be sure, this particular issue may be definitively resolved 
only after a careful study of the corresponding vertex SDE and the Bethe-Salpeter equation derived from it
(see the related discussion and references in Sec.~\ref{sec:num}).

Since, apart from the qualitative considerations given above, 
the precise form of the function $\widetilde{C}_1^{\prime}(y)$ is undetermined at this level, in order to proceed with our analysis we consider three possible models describing its functional form. The general shapes chosen are inspired by
the solutions obtained from the Bethe-Salpeter equations governing the formation of
massless poles~\cite{Aguilar:2011xe}. In particular, we use the following models
\begin{align}
	\widetilde{C}_1^{\prime}(y)=
	\begin{cases}
    1/(ay^2 + by+ c), & \text{Model A } \\
    ay\exp(-y/b), & \text{Model B }  \\
    1/(ay + b\sqrt y+ c), & \text{Model C }
  \end{cases}
  \label{Cprimemodels}
\end{align}
with $a$, $b$, and $c$ suitable parameters. 

\begin{figure}[!t]
\includegraphics[scale=.615]{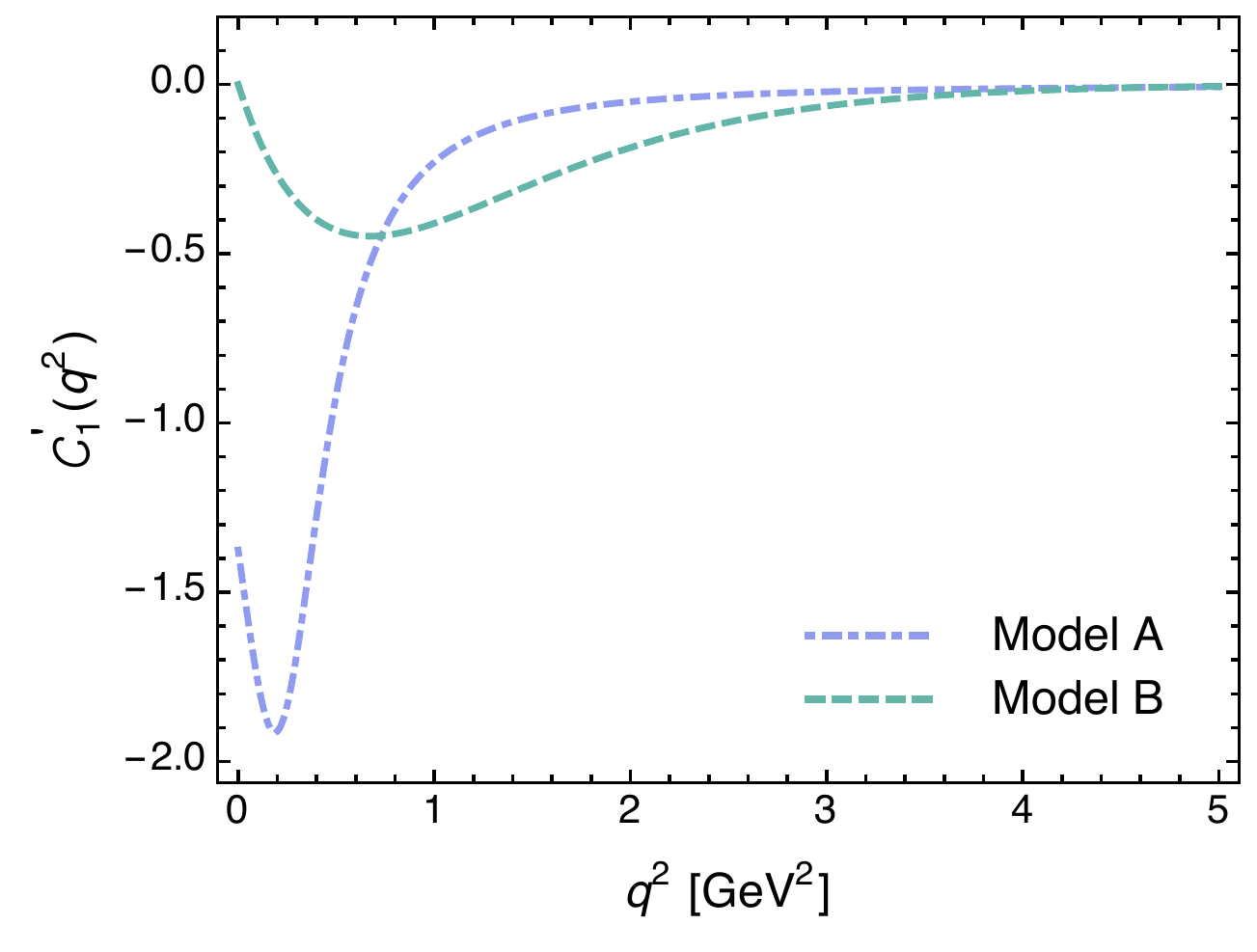}
\includegraphics[scale=.615]{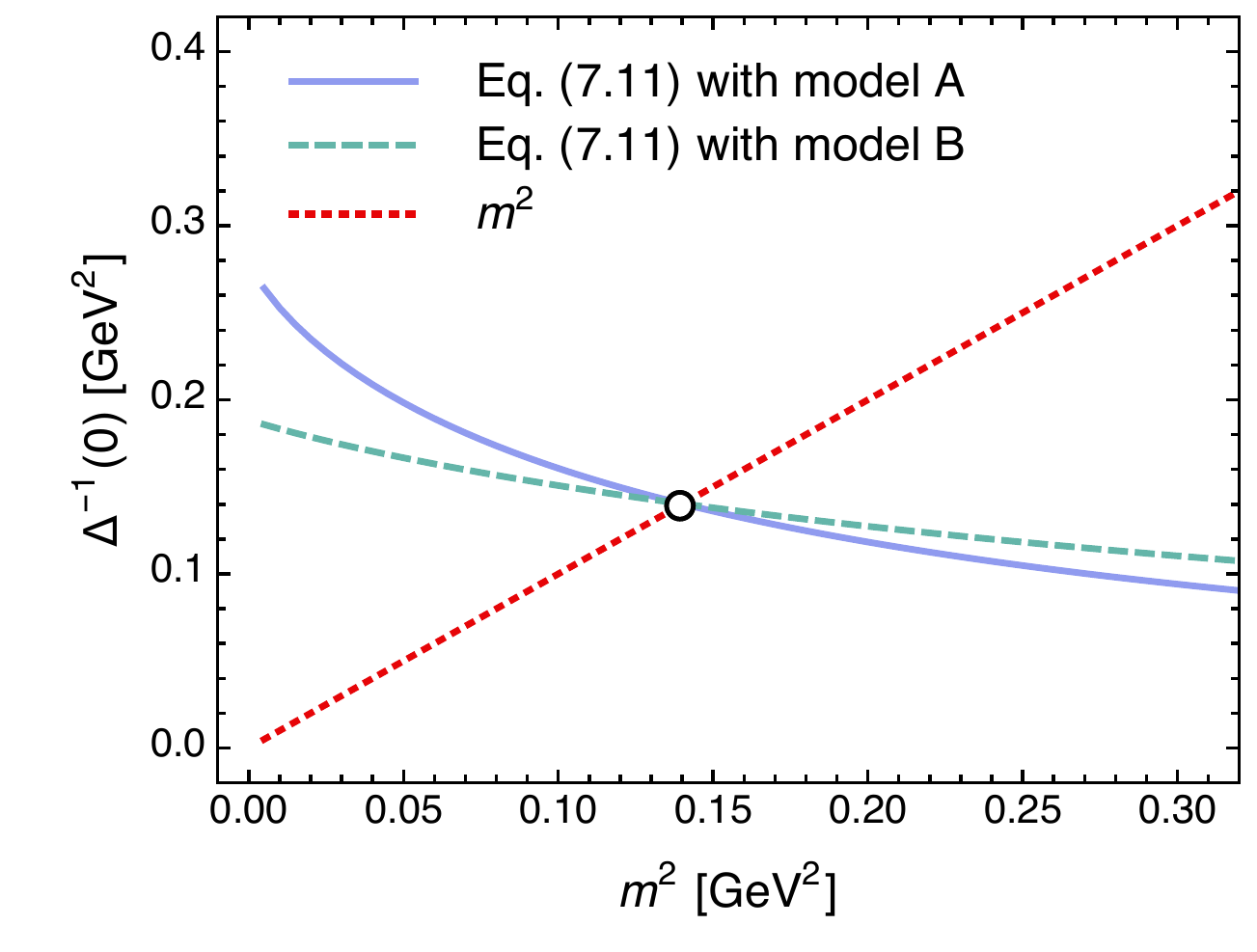}
\caption{\label{fig:Massive-model}(Color online) Left panel: The functions $\widetilde{C}_1^{\prime}(y)$ obtained from 
Model A with \mbox{$a = -5.82 \,\mbox{GeV}^{-4}$}, $b=2.18\, \mbox{GeV}^{-2}$, $c=-0.73$, and from 
Model B with $a= -1.81\,\mbox{GeV}^{-2}$, $b=0.675 \,\mbox{GeV}^{2}$. Right panel: 
The solution of \1eq{num5a} when $m^2$ coincides with the SU$(3)$ lattice saturation value $\Delta^{-1}(0)=0.14$ GeV$^2$.}
\end{figure}

To begin with let us assume that the gluon propagator has a simple massive form, \mbox{$\Delta(y)=1/(y+m^2)$}, so that~\1eq{num5} becomes
\begin{align}
	\Delta^{-1}(0)=m^2=-\frac{3 C_A \alpha_s}{8\pi} F(0)\int_0^{\infty}\! \mathrm{d}y\, \frac{y^2}{(y+m^2)^2}\widetilde{C}_1^{\prime}(y).
	\label{num5a}
\end{align}
To proceed further, we will use as input in \1eq{num5a} the values
for the saturation points of the  gluon propagator and the ghost dressing function found in the  
lattice simulations of \cite{Bogolubsky:2009dc,Bogolubsky:2007ud}; specifically, 
when the MOM subtraction point is chosen at $\mu=4.3$ GeV, one has that $\Delta^{-1}(0)= m^2=0.15$ GeV$^2$ and $F(0)=2.91$.
The idea then is to try to determine the parameters of $\widetilde{C}_1^{\prime}(y)$ in \1eq{Cprimemodels} so that \1eq{num5a} is satisfied. 
Of course, the solution of this problem is not unique, as there are many possible  
combinations of $a$, $b$, and $c$ leading to the same result. In addition, notice that, for 
the case of a simple massive propagator, model C cannot be considered, 
because the corresponding integral diverges logarithmically at its upper limit.
In \fig{fig:Massive-model} we show how the desired solution is obtained using the particular set of values 
for the parameters of models A and B given in the caption.

\begin{figure}[!t]
\hspace{-.5cm}
\includegraphics[scale=.615]{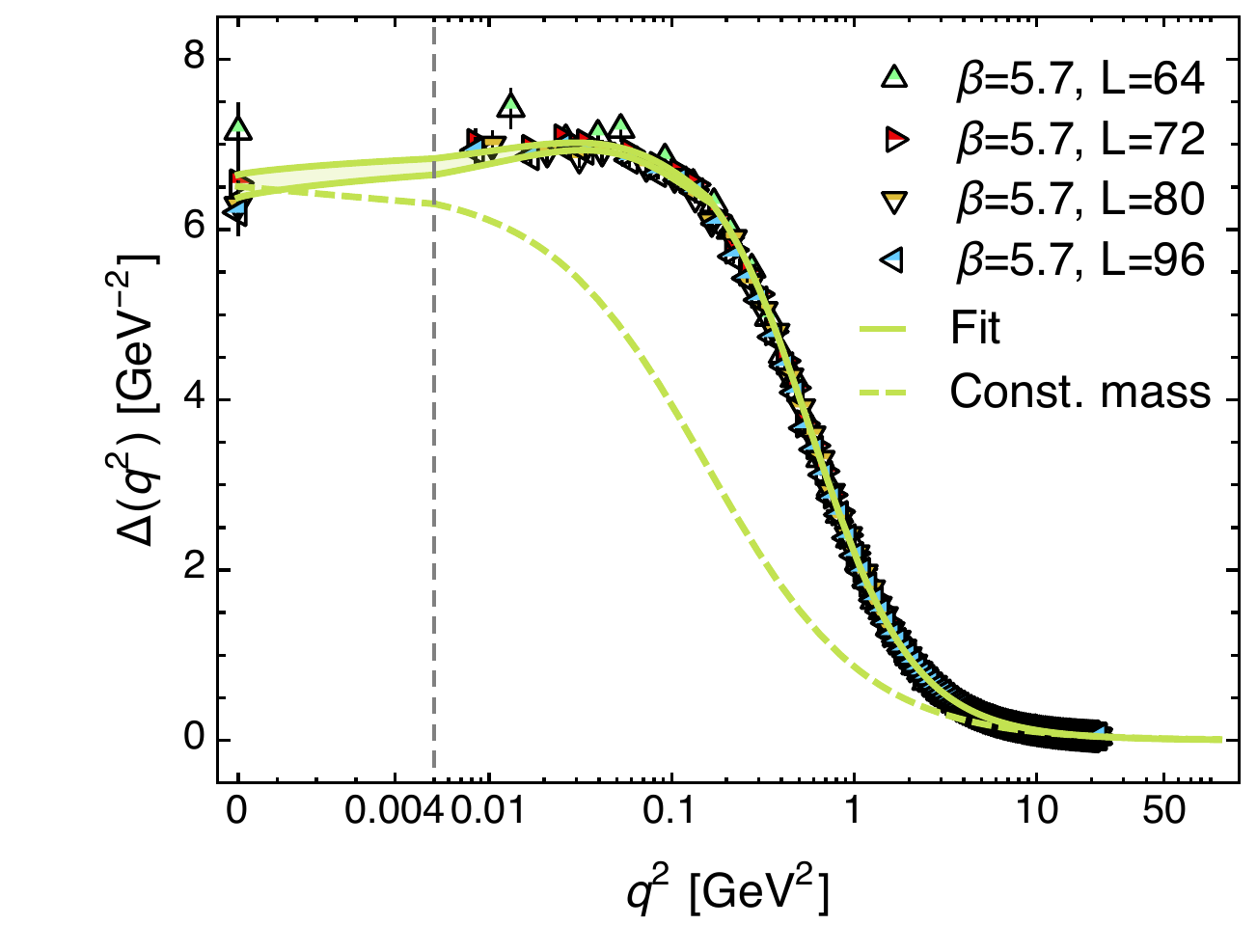}
\includegraphics[scale=.615]{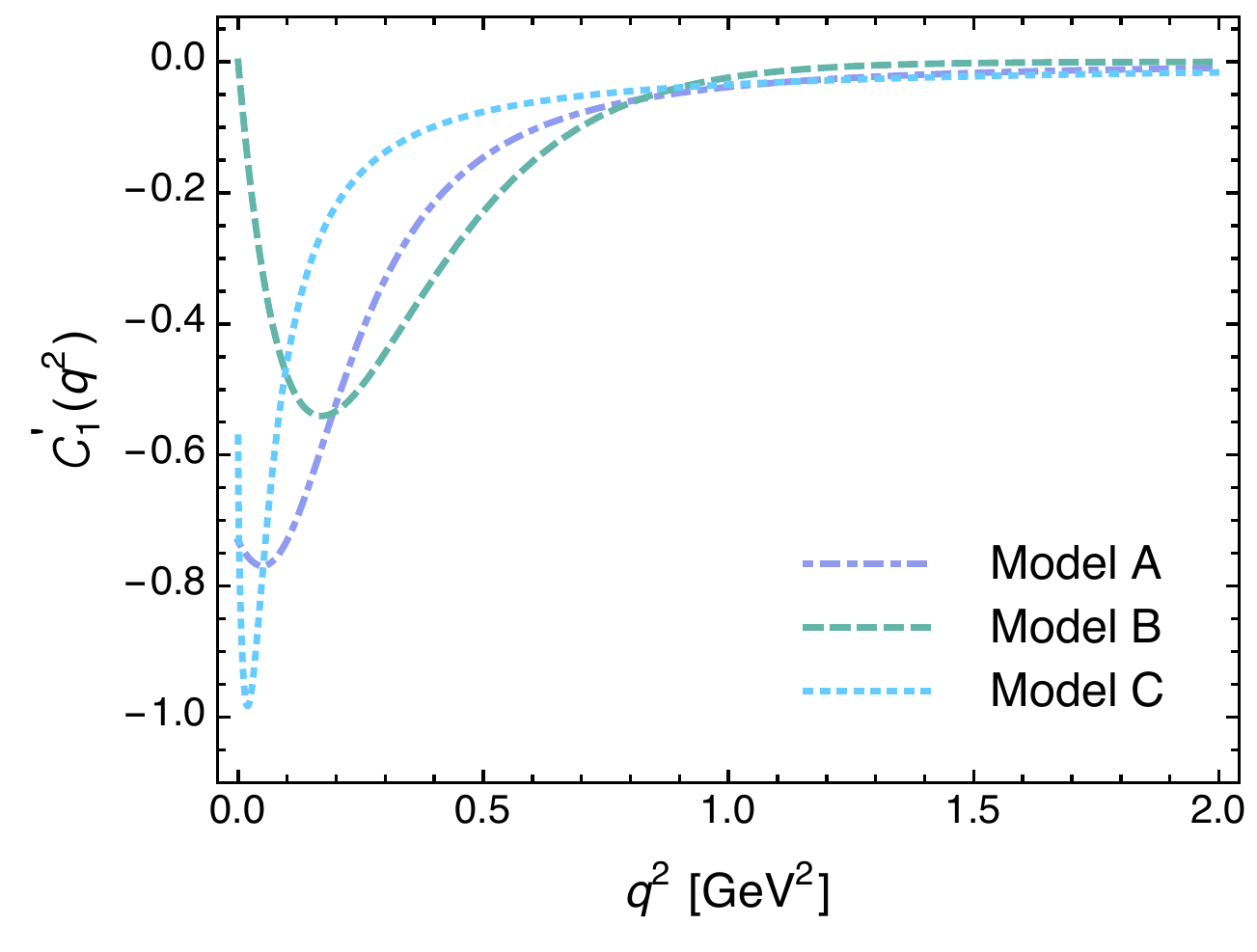}
\caption{\label{fig:Lattice-model}(Color online) Left panel: The quenched SU$(3)$ lattice data for the gluon propagator, $\Delta(q^2)$, renormalized at \mbox{$\mu=4.3$ GeV} (triangles) and its corresponding fit (continuous line); for comparison, we also plot a simple constant mass propagator (dashed line). Notice that the scale on the $x$-axis becomes linear past the dashed vertical line.   
Right panel: 
The functions $\widetilde{C}_1^{\prime}(y)$ correspond to:  
Model A with $a = -26.78 \,\mbox{GeV}^{-4}$, $b=2.73\, \mbox{GeV}^{-2}$, $c=-1.37$, 
Model B with $a= -8.64\,\mbox{GeV}^{-2}$, $b=0.17 \,\mbox{GeV}^{2}$, and 
Model C with $a = -37.2 \,\mbox{GeV}^{-2}$, $b=-10.44 \,\mbox{GeV}^{-1}$, $c=-1.74$; all three curves yield a solution to~\1eq{num5} when the lattice gluon propagator is used as input.} 
\end{figure}

Let us next repeat the  above analysis employing a more realistic gluon propagator, namely the one obtained in the SU(3) quenched lattice simulations of Ref.~\cite{Bogolubsky:2007ud}. On the left panel of \fig{fig:Lattice-model} we show a physically motivated fit for the gluon propagator (green continuous line), together with the lattice data at the renormalization scale $\mu=4.3$ GeV (triangles). Notice that the fit displays the inflection point that must appear due to the presence of divergent ghost loops~\cite{Aguilar:2013vaa}; we will comment on this issue shortly.

We clearly see that the lattice gluon propagator is significantly more enhanced in the region below \mbox{$2\,\mbox{GeV}^2$} when compared to the naive constant mass propagator used in our previous discussion, with $m^2=0.15$ GeV$^2$ (dashed curve). It is therefore interesting to study how the  values of the model parameters $a$, $b$, and $c$ must be modified in order to obtain from \1eq{num5} the same solution as before.

A typical case is shown on the right panel of \fig{fig:Lattice-model}, where we 
plot the $\widetilde{\cal C}'(y)$ obtained for each model considered, for the parameters quoted in the caption. 
Evidently, the $\widetilde{C}_1^{\prime}(y)$ used in this case are suppressed compared to those of the simple massive case,
because, precisely due to the aforementioned enhancement of the gluon propagator, 
less strength is required from them in order for the r.h.s. of \1eq{num5} to reach the
fixed value of $0.15$~GeV$^2$.

We end this section with a brief qualitative description of how the study of 
the form factor $\widetilde{A}^{\mathbf{np}}_6(q^2)$, and in particular of the relation \1eq{A6np}, 
may corroborate or invalidate the necessity of longitudinally-coupled massless poles.
This, in turn, may be particularly interesting, especially in view of the comments following \1eq{Aipnp}.

\begin{figure}[!t]
\includegraphics[scale=0.7]{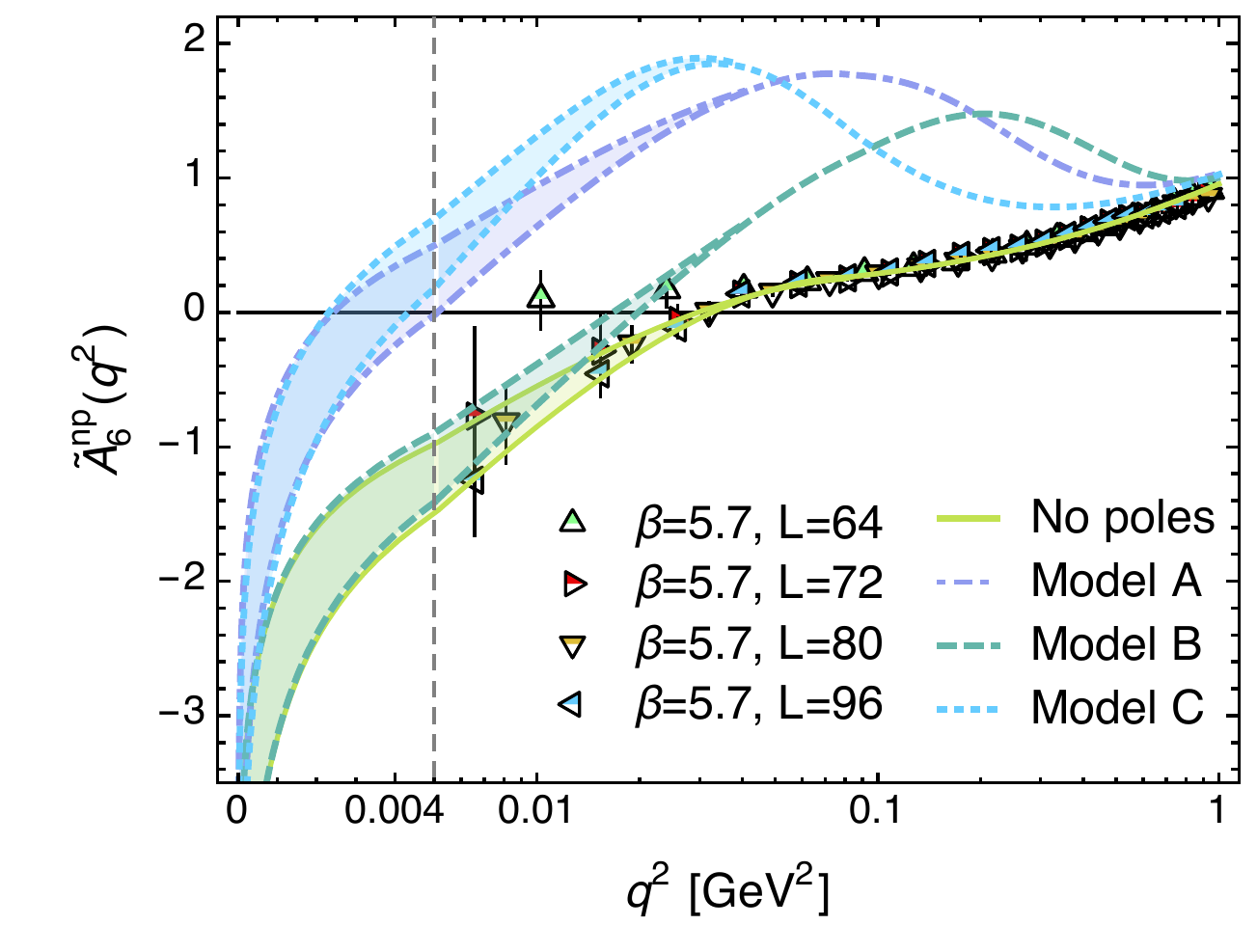}
\caption{\label{fig:qA6} (Color online) The form factor $\widetilde{A}^{\bf{np}}_6$ evaluated when using the quenched SU(3) lattice propagator as input. In the absence of massless poles, this quantity is proportional to the derivative of the inverse propagator (here evaluated directly from the lattice data), see~\1eq{theA}; notice the zero crossing and logarithmic divergence in the deep infrared (again past the vertical dashed line the scale becomes linear). The band indicates the spreading between the $L=72$ and $L=96$ lattice data. When massless poles are generated, the presence of the function $\widetilde{C}_1^{\prime}$ modifies this dependence according to~\1eq{A6np}, and a positive maximum appears in the region $q^2<1$ GeV$^2$, the height and exact location of which depend on the details of the model.}
\end{figure}

To begin with, let us recall that, in contradistinction to the  `massive' behavior observed for the gluon propagator, the Landau gauge lattice simulations reveal that the ghost remains massless: the ghost propagator behaves like $1/p^2$ in the IR, and is multiplied by a dressing function that saturates at a finite nonvanishing value~\cite{Cucchieri:2007md,Cucchieri:2007rg,Cucchieri:2009zt,Cucchieri:2010xr,Bogolubsky:2009dc,Bogolubsky:2007ud}.
Combining these two results together, one arrives at the (model-independent) conclusions that~\cite{Aguilar:2013vaa} 
\n{i} the gluon propagator must display a mild maximum in the deep infrared (a feature that can be seen in the left panel of \fig{fig:Lattice-model}), 
and \n{ii} the derivative of its inverse will display a 
logarithmic divergence (and a corresponding zero crossing). In particular, one has that    
\begin{align}
	\frac{\partial}{\partial q^2}\Delta^{-1}(q^2)\underset{q\to0}{\sim}\log q^2,
	\label{dataderiv}
\end{align}
and indeed, by evaluating the derivative of the lattice data, one can see the appearance of the zero crossing and 
the logarithmic divergence predicted by~\1eq{dataderiv} (see data points in \fig{fig:qA6}).  

Thus, if the observed finiteness of the gluon propagator were not due to the presence of massless poles, 
\1eq{theA} implies that the three-gluon vertex form factor $\widetilde{A}^{\bf{np}}_6\equiv\widetilde{A}_6$ 
would display the same infrared behavior as that of \1eq{dataderiv}. On the other hand, the presence of 
massless poles will  alter the shape of this form factor, as the function $\widetilde{C}_1^{\prime}$ 
will now enter in its determination [see~\1eq{A6np}]. Given the finiteness of $\widetilde{C}_1^{\prime}$, 
this will not alter the deep infrared behavior of the form factor; however, it will force it 
to have a characteristic positive maximum, which could be detected against the dominant term, 
as shown in \fig{fig:qA6}. In fact, the best strategy might be to determine 
simultaneously $\Delta$ and $\widetilde{A}_6$ and form 
the difference $\widetilde{A}_6-2\partial\Delta^{-1}$, which would yield directly the function ${\cal C}'_1$ 
(multiplied by a factor of 2).

\section{\label{sec:c}Discussion and Conclusions}                                                        

We  have  presented  a  unified   framework  for  the self-consistent treatment of the infrared finiteness of the gluon propagator at the  level of  the SDEs  of the  theory, formulated within the PT-BFM framework. Particular attention has been dedicated to the extensive cancellations induced by the WIs of the theory, and the necessity of introducing massless poles in order to achieve the desired effect in a self-consistent way. Our analysis has been carried out for a general value of the gauge-fixing parameter, reverting to the Landau gauge only in order to simplify the numerical analysis presented in the last section.

Particularly relevant is the observation that the presence of these poles is bound to affect not only the two-point but also the 
three-point sector of the theory. Therefore, determining particular form factors of, {\it e.g.}, the (background) three-gluon vertex by means of lattice techniques, has the potential 
to confirm or discard massless poles as the infrared mechanism underlying the dynamical generation of a gluon `mass'. The main difficulty 
in actually realizing this idea 
stems from the fact that the form factors satisfying~\2eqs{Awinp}{A6np} are those of 
the $BQ^2$ and $Bc\bar c$ vertices, rather than those comprising the conventional vertices $Q^3$ and $Qc\bar c$~\cite{Huber:2012kd,Blum:2014gna,Eichmann:2014xya}. 
The two sets of vertices are related through rather complicated `background-quantum identities~\cite{Binosi:2008qk}, 
which need to be studied in order to determine if any simplifications occur in the relevant $q\to0$ limit. At the same time, it might be 
also helpful to explore this issue from the lattice point of view~\cite{Cucchieri:2006tf,Cucchieri:2008qm}, by implementing the BFM 
along the preliminary proposals put forth in~\cite{Binosi:2012st,Cucchieri:2012ii}. 

It is clear from the analysis presented that the three-gluon vertex, 
and in particular its PT-BFM version, $BQ^2$, 
 plays a particularly important role in the 
entire construction that leads to an infrared finite gluon propagator. One of the most notable features 
is the presence of massless poles in a special subset of its form factors, 
which are therefore expected to diverge in the deep infrared. It is important to mention that 
several SDE studies of the gluon and ghost propagator, as well as of higher order Green's functions, have considered  
the possibility that the conventional three-gluon vertex, $Q^3$, displays a divergent behavior.
Some of the most representative works in this direction include: 
\n{i} Focusing only on the form factors of the tensors appearing in the tree-level vertex [first equation in (\ref{A2})], 
and using  a power-counting scheme, the authors of~\cite{Alkofer:2004it}  
found that the infrared limit of the ``symmetric configuration'' 
($q^2=r^2=p^2$) shows a divergent power-law behavior of the type $(p^2)^{-3\kappa}$, 
where $\kappa=0.59$ is the typical parameter of the so-called ``scaling solutions''.
\n{ii}  A similar analysis was
performed in~\cite{Alkofer:2008jy}, for one soft and two hard external momenta, finding  
a softer divergence, of the type $(p^2)^{1-2\kappa}$. 
\n{iii} A study with 
a more complete tensorial structure was carried out in~\cite{Alkofer:2008dt}, where 
the {\it transverse} parts of the three-gluon vertex turned out to be very mildly divergent, and with  
no appreciable impact on the gluon and ghost SDEs. 
We would like to emphasize that, even though the main qualitative features between the aforementioned results and those of the present work 
appear to be similar,  a direct quantitative comparison between them is   
not possible at present, mainly due to the fact that the fundamental properties of the 
$Q^3$ and $BQ^2$ vertices are very different. In fact, as mentioned in the previous paragraph, the 
two vertices are formally related by an exact, but rather complicated identity~\cite{Binosi:2008qk}, 
the ingredients of which are, to a large extent, unexplored. 
 

The actual generation of the poles as massless bound-state excitations
may be studied in the generalized context of the Bethe-Salpeter equations, as proposed in the early works of~\cite{Jackiw:1973tr,Jackiw:1973ha,Eichten:1974et,Poggio:1974qs}, and  as was further explored in~\cite{Aguilar:2011xe,Ibanez:2012zk}. Note, however, that in the analysis presented in~\cite{Aguilar:2011xe,Ibanez:2012zk} the possibility that the ghost-gluon vertex may contain such poles was not contemplated, and $\gp_{\mu}$ was assumed to vanish identically. It is therefore especially interesting to explore whether or not the formation of a nontrivial $\gp_{\mu}$ is dynamically favored. In particular, whereas in previous considerations only the Bethe-Salpeter equation for the massless pole of the three-gluon vertex was studied, under the light of the analysis presented the corresponding dynamical equation for the ghost vertex must be derived and solved. In fact, the complete treatment of this problem would require the solution of a coupled system rather than a single integral equation, given that the pole 
part of the ghost vertex gets mixed with that of the three-gluon vertex, and vice-versa.  

The general formalism developed in this work sets up the stage for a detailed quantitative study of the precise field-theoretic mechanism that accounts for the infrared saturation of the gluon propagator observed in recent lattice simulations performed away from the Landau gauge~\cite{Bicudo:2015rma}. In particular, the relevant set of Bethe-Salpeter equations must be derived for general $\xi$, and then appropriately coupled to \3eqs{0a1a2mass}{0a3a4mass}{0a5a6mass}, which determine the value of $\Delta^{-1}(0)$. This analysis may be particularly revealing, given the observed tendency of the gluon saturation point to decrease as $\xi$ increases, at least within the interval $[0,0.5]$. Reproducing this characteristic behavior constitutes a considerable challenge, whose successful completion would considerably validate the proposed approach and general philosophy. We hope to be able to pursue this issue in the near future.
  
\acknowledgments 

The research of J.~P. is supported by the Spanish MEYC under grants FPA2014-53631-C2-1-P and SEV-2014-0398, and Generalitat Valenciana  
under grant Prometeo~II/2014/066. The work of  A.~C.~A  is supported by the National Council for Scientific and Technological Development - CNPq under the grant 305815/2015. C.~T.~F. acknowledges the financial support from S\~ao Paulo Research Foundation - FAPESP through the project 2014/16247-8. 
We thank D.~Iba{\~n}ez for his contribution at an early stage of this work.

\appendix
\section{\label{app:A}Feynman rules}

The following vertex definitions have been employed (all momenta entering): 
\begin{align}
	&i\Gamma_{Q^a_\mu Q^m_\alpha Q^n_\beta}(q,r,p)=gf^{amn}\Gamma_{\mu\alpha\beta}(q,r,p);&
	&i\Gamma_{B^a_\mu Q^m_\alpha Q^n_\beta}(q,r,p)=gf^{amn}\widetilde{\Gamma}_{\mu\alpha\beta}(q,r,p),&\nonumber \\
	&i\Gamma_{c^nQ^a_\mu\bar c^m}(p,q,r)=gf^{amn}\Gamma_\mu(q,r,p);&
	&i\Gamma_{c^nB^a_\mu\bar c^m}(p,q,r)=gf^{amn}\widetilde{\Gamma}_\mu(q,r,p);& \nonumber \\
	&\Gamma_{Q^a_\mu Q^m_\alpha Q^n_\beta Q^r_\gamma}(q,r,p,t)=-ig^2\Gamma^{amnr}_{\mu\alpha\beta\gamma}(q,r,p,t);&
	&\Gamma_{B^a_\mu Q^m_\alpha Q^n_\beta Q^r_\gamma}(q,r,p,t)=-ig^2\widetilde{\Gamma}^{amnr}_{\mu\alpha\beta\gamma}(q,r,p,t).&
\end{align}
At tree-level one has
\begin{align}
	& \Gamma^{(0)}_{\mu\alpha\beta}(q,r,p)=g_{\alpha\beta}(r-p)_\mu+g_{\mu\beta}(p-q)_\alpha+g_{\mu\alpha}(q-r)_\beta,\nonumber \\
	& \widetilde{\Gamma}^{(0)}_{\mu\alpha\beta}(q,r,p)=g_{\alpha\beta}(r-p)_\mu+g_{\mu\beta}(p-q+\xi^{-1}r)_\alpha+g_{\mu\alpha}(q-r-\xi^{-1}p)_\beta,\nonumber \\
	& \Gamma^{(0)}_\mu(q,r,p)=-r_\mu,\nonumber \\
	& \widetilde{\Gamma}^{(0)}_{\mu}(q,r,p)=(p-r)_\mu, \nonumber \\
	& \Gamma^{(0)\,amnr}_{\mu\alpha\beta\gamma}(q,r,p,t)=f^{are}f^{enm}(g_{\mu\beta}g_{\alpha\gamma}-g_{\mu\alpha}g_{\beta\gamma})+f^{ame}f^{ern}(g_{\mu\gamma}g_{\alpha\beta}-g_{\mu\beta}g_{\alpha\gamma})\nonumber\\
	&\hspace{3.1cm}+f^{ane}f^{erm}(g_{\mu\gamma}g_{\alpha\beta}-g_{\mu\alpha}g_{\beta\gamma}),\nonumber \\
	&\widetilde{\Gamma}^{(0)\,amnr}_{\mu\alpha\beta\gamma}(q,r,p,t)=\Gamma^{(0)\,amnr}_{\mu\alpha\beta\gamma}(q,r,p,t). 
\label{A2}
\end{align}


\end{document}